\documentclass[nohyper,notoc]{JHEP3}

\usepackage{amsmath,euscript,array,amssymb,cite,bm} 

\usepackage{graphicx}
\usepackage{subfigure}
\def\yp{{(Y')}}

\def\Tr{{\rm Tr}}

\def\Dbarslash{\,\,{\raise.15ex\hbox{/}\mkern-12mu {\bar\D}}}
\def\Dslash{\,\,{\raise.15ex\hbox{/}\mkern-12mu \D}}
\def\delslash{\,\,{\raise.15ex\hbox{/}\mkern-9mu \partial}}
\def\delbarslash{\,\,{\raise.15ex\hbox{/}\mkern-9mu {\bar\partial}}}

\def\D{{\cal D}}
\def\Dbarslash{\,\,{\raise.15ex\hbox{/}\mkern-12mu {\bar\D}}}
\def\delslash{\,\,{\raise.15ex\hbox{/}\mkern-9mu \partial}}
\def\Dslash{\,\,{\raise.15ex\hbox{/}\mkern-12mu \D}}

\newcommand{\be}{\begin{equation}}
\newcommand{\ee}{\end{equation}}
\def\bea{\begin{eqnarray}}
\def\eea{\end{eqnarray}}

\newcommand{\ialpha}{\alpha^{-1}}

\def\uno{\mbox{1 \kern-.59em {\rm l}}}

\title{Direct Mediation, Duality and Unification}

\author{Steven Abel and Valentin V.~Khoze\\

Institute for Particle Physics Phenomenology, University of Durham,
Durham, DH1 3LE, UK

{\tt s.a.abel@durham.ac.uk,}\,
{\tt valya.khoze@durham.ac.uk}}

\abstract{It is well-known that in scenarios with {\it direct} gauge mediation of supersymmetry breaking 
the messenger fields significantly affect the running of Standard Model couplings and introduce 
Landau poles which are difficult to avoid. Among other things, this
appears to remove any possibility of a meaningful unification prediction and is often viewed as a strong argument against direct mediation.
We propose two ways that Seiberg duality can circumvent this problem. In the 
first, which we call ``deflected-unification'', 
the SUSY-breaking hidden sector is a magnetic theory 
which undergoes a Seiberg duality to an electric phase. Importantly, the electric 
version has fewer fundamental degrees of freedom coupled to the MSSM compared to the magnetic formulation. This
changes the $\beta$-functions of the MSSM gauge couplings so as to 
push their Landau poles above the unification scale. We 
show that this scenario is realised for 
recently suggested models of gauge mediation based on a metastable SCQD-type  hidden sector directly coupled to MSSM. The second possibility for avoiding Landau poles,
which we call ``dual-unification'', begins with the observation 
that, if the mediating fields fall into complete $SU(5)$ multiplets, then 
the MSSM+messengers exhibits a fake unification at unphysical values of 
the gauge couplings. We show that, in known examples of electric/magnetic 
duals, such a fake unification in the magnetic theory reflects a 
real unification in the electric theory. We therefore propose that the Standard Model could itself be a magnetic dual
of some unknown electric theory in which the true unification takes place.
This scenario maintains the unification prediction
(and unification scale) even in the presence of Landau poles in the magnetic theory below the 
GUT scale. We further note that this dual realization of grand unification can explain why 
Nature appears to unify, but the proton does not decay.}

\preprint{IPPP/08/72\\
DCPT/08/144
}

\begin{document}

\section{Introduction}

One of the key questions of supersymmetric BSM particle physics is the nature of 
supersymmetry breaking in the hidden sector and its mediation to the visible Standard Model sector.
A particularly concise and appealing proposal is the {\it direct} gauge mediation approach \cite{Affleck:1984xz}
in which the supersymmetry-breaking sector couples directly to the MSSM with no need for an additional messenger sector.

The central issue we wish to address in this paper 
applies particularly to models of direct gauge mediation and is this.
In direct gauge mediation, supersymmetry (SUSY) is broken
in a hidden sector that contains a large 
global flavour group which is gauged and identified with the 
gauge group of the Supersymmetric Standard Model (SSM) or a subgroup thereof. 
Supersymmetry breaking is mediated by
messengers that are charged under both 
groups~\cite{Affleck:1984xz,Luty:1998vr}. Being charged under the hidden sector gauge symmetry,
these direct messengers are in effect a large number of additional matter fields 
for the visible sector\footnote{In general we will refer to the supersymmetric Standard Model including direct messengers as the SSM. We will reserve "MSSM" to refer to the minimal model on its own (without messengers).}
. Consequently they give a large positive contribution to the 
$\beta$-functions above the messenger scale, $M_{mess}$, and
the visible sector gauge couplings then encounter Landau poles below
$M_{GUT}$ at which point the
perturbative description breaks down. Thus to use direct gauge mediation
one has apparently to abandon
perhaps the most successful prediction of the MSSM, 
namely gauge unification. This is often cited as evidence against it. 

Here we wish to propose that theories with 
Landau poles can still have meaningful and predictive unification  
taking place ``across a Seiberg duality'' in a microscopic electric dual 
description.\footnote{Other mechanisms of avoiding Landau poles have also been considered in the literature, including
\cite{Poppitz:1996fw,ArkaniHamed:1997jv,Murayama:1997pb} and most recently in \cite{Seiberg:2008qj}.}
The Seiberg duality can be applied either to just the 
hidden sector or may also include the visible sector. In the former case the 
universal change of slopes provided by messengers in complete 
$SU(5)$ multiplets persists, but the slopes can change a number of 
times as the hidden sector goes through one or more dualities. 
This is a form of ``deflected-unification'' (to borrow a term from 
anomaly mediation) as shown schematically in figure 1. Similar
deflection of course happens in purely perturbative theories
whenever a threshold is crossed, but
this always results in an increase in the effective number of flavours which
only accelerates the running to Landau poles. 
The important feature that the Seiberg duality \cite{Seiberg} brings is a 
{\em reduction} in the number of elementary direct messenger fields
when one switches from the IR magnetic to the UV electric formulation of the hidden sector theory.
This reduction affects the $\beta$-function slopes in the SSM and moves the Landau poles to the UV or 
conceivably even removes them entirely. Note that the 
slope in the visible sector changes in this way only because the mediation 
is direct; from the point of view of the visible sector the electric theory
simply has a different number of messenger flavours contributing 
in the one-loop $\beta$-functions. 

What solves the problem of Landau poles in this class of direct mediation scenarios is the 
assumption that the SUSY-breaking sector has an electric dual in the UV. 
Remarkably, this is precisely what happens in the Intriligator, Seiberg and Shih (ISS) model \cite{iss}.
In the ISS case Seiberg duality was instrumental in achieving dynamical SUSY-breaking (DSB) in a metastable vacuum. Now, when the ISS-type model is used as the hidden sector for direct mediation of SUSY-breaking to the SSM
\cite{Csaki,Kitano,us,us2,oz}, it not only provides us with a simple satisfactory description of DSB, it also resolves
the Landau pole problem of direct mediation. In Section 2 we will show how this works in the context of 
the direct mediation models introduced in \cite{us,us2}.

\begin{figure}[h]
\begin{centering}
\includegraphics[bb=0bp 40bp 450bp 450bp,clip,angle=0,scale=0.4]{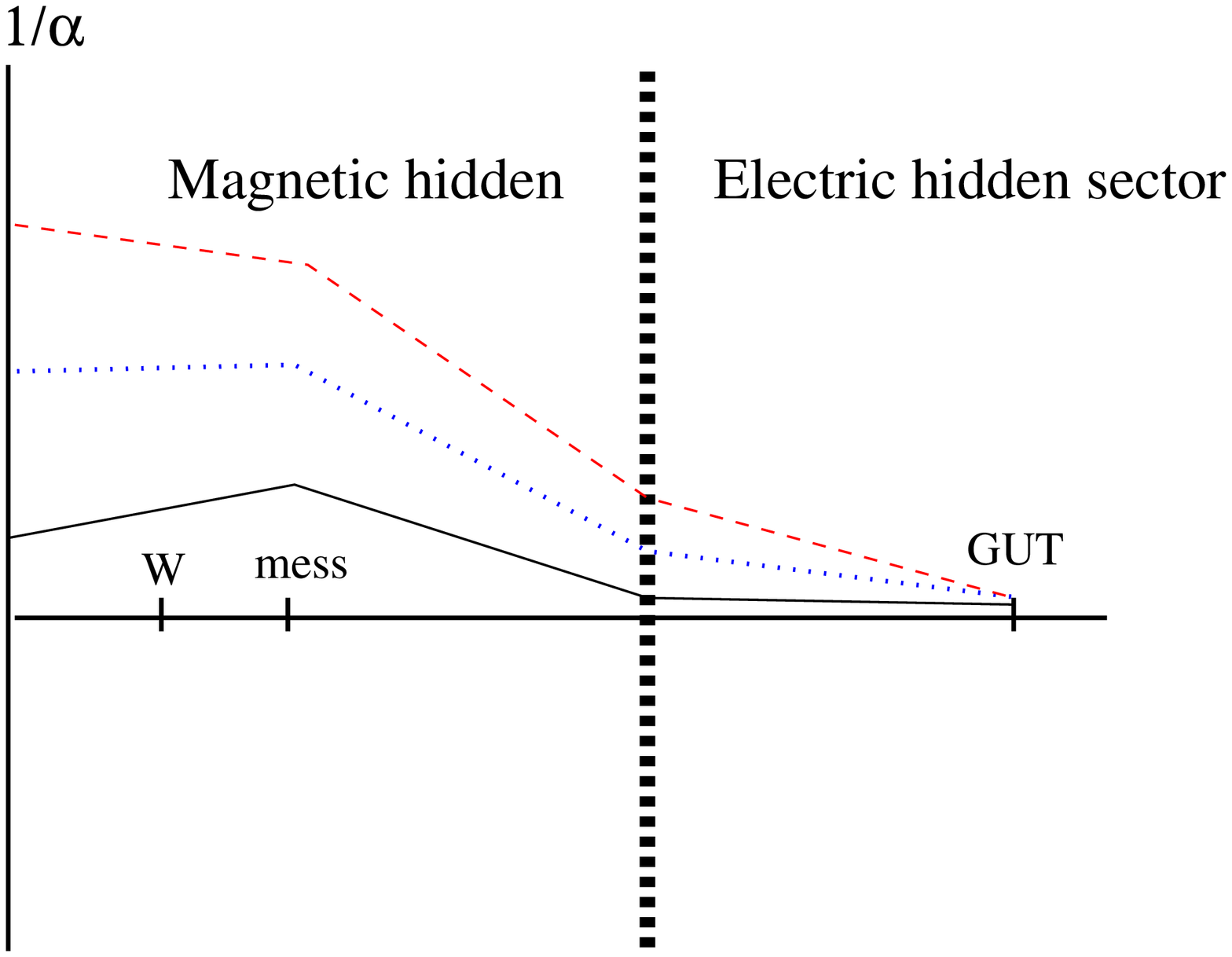}
\par\end{centering}
\caption{\it Schematic set-up for deflected-unification. The SSM couplings ($U(1)_Y\equiv\mbox{red/dashed}$;
$SU(2)\equiv\mbox{blue/dotted}$; $SU(3)\equiv\mbox{black/solid}$)
experience accelerated running above a messenger scale, but their 
running is deflected again where the magnetic hidden sector theory is 
matched to the electric one.}
\end{figure}

The complementary possibility is that the duality takes place in the 
visible sector as well. One hint that unification may be preserved under such a 
duality has been noted by several authors in this context
including most recently in ref.\cite{oz}, 
 and is the following: 
in a model of direct gauge mediation with
messengers in complete $SU(5)$ multiplets the extra contributions to  
$\beta$-functions are degenerate; they run to strong coupling 
well below $M_{GUT}$, however the relative running is unchanged,
so the three gauge couplings of the SSM still appear to unify at
the scale $M_{GUT}$, but at 
unphysical, negative values of $\alpha_{i=1,2,3}=\alpha_{GUT}$. Could this fake unification be a remnant of a real unification in the electric theory?

\begin{figure}[h]
\begin{centering}
\includegraphics[bb=0bp 40bp 450bp 450bp,clip,angle=0,scale=0.4]{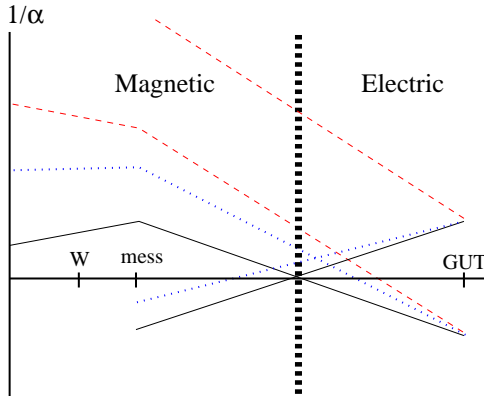}
\par\end{centering}
\caption{\it Schematic set-up for dual-unification. The SSM couplings ($U(1)_Y\equiv\mbox{red/dashed}$;
$SU(2)\equiv\mbox{blue/dotted}$; $SU(3)\equiv\mbox{black/solid}$)
experience accelerated running above a messenger scale, but still
appear to unify at unphysical values. 
The unification takes place at physical values in the electric dual theory. 
Note that the $U(1)_Y$ is rescaled in the dual magnetic theory rather than 
dualized, and thus its slope does not change unless the number of quark 
flavours changes.}
\end{figure}
In this paper (Sections 3-4) we will see that indeed it can be: unification in the
electric description leads to a {}``fake'' unification in the magnetic
one at the same energy scale but at unphysical (i.e. negative) values of $\alpha^{-1}_{i}$.
This occurs where the entire unified electric theory is
dualized (i.e. if we are thinking of the SSM, then every subgroup
of the SSM gets dualized). The generic picture, which we call 
``dual-unification'',
is as sketched in figure 2. We show that a large class of known
electric/magnetic duals exhibit dual-unification, and that
under very general assumptions it is guaranteed by the matching relations 
between the electric and magnetic theories.

Finally, in Section 5 we argue that the dual-unification scenario has significant 
implications for the question of proton decay. While it is well-known that in the usual simple MSSM-GUT 
picture the lifetime of the proton turns out to be shorter than experimental bounds, we will
show that in the dual-unification set-up proton decay processes can be enormously suppressed. 
This is due to the fact that the dangerous baryon number violating 
operators are induced in the electric theory where the unification takes place.
At this energy scale the magnetic theory is strongly coupled and one must instead use the 
weakly coupled electric theory description, and then map to the low energy magnetic theory with well 
know baryon$_{mag}\leftrightarrow $\,\,baryon$_{elec}$ identifications. 
This introduces many powers of $\Lambda /M_{GUT} \ll 1$ making proton decay completely negligible.

\section{Deflected-unification}

One of the most interesting and appealing properties of supersymmetric
theories is the holomorphy and nonrenormalizability of their superpotentials;
from these two properties many powerful statements follow about the
nonperturbative effects of strong coupling. In particular in a large
number of celebrated examples beginning with $\mathcal{N}=1$ SQCD
models\cite{Seiberg}, one can find two (or more) dual theories that
describe the same IR physics (for a review see refs.\cite{Intriligator:1995au,Terning:2003th}).
For certain choices of parameters, a theory can enjoy two perturbative
regimes, an asymptotically free electric one that accurately describes
the UV physics and a free magnetic phase that describes the IR physics.
This is the situation that will be of interest for this paper. 

Let us see how such electric/magnetic duality can effect unification in
simple direct mediation. 
We will consider a theory in which a hidden sector couples directly to the
visible sector through bifundamental fields, charged under both the 
visible and hidden gauge groups. As described in the Introduction, 
if the hidden
sector gauge group undergoes a duality at some energy scale, then
the number of flavours seen by the visible
sector also changes at that scale. If the multiplets coupling hidden
to visible sector are in complete $SU(5)$'s as is often assumed in
gauge mediation, then the nett effect will be a universal change of slope
which can allow unification where in the magnetic theory it appears
to be impossible. 

A simple example is provided by the model of refs.\cite{us,us2}.
This is a model of direct gauge mediation with a supersymmetry breaking
ISS sector \cite{iss} with $N_{f}=7$, and $N_{c}=5$. This sector
becomes strongly coupled at a scale $\Lambda_{ISS}$, and can be described
by an IR free magnetic $SU(2)_{mg}$ theory below that scale. The relevant 
quiver diagrams are shown in figure 3. 
\begin{figure}[b]
\begin{centering}
\includegraphics[scale=0.8, bb=15bp 560bp 700bp 790bp,clip]{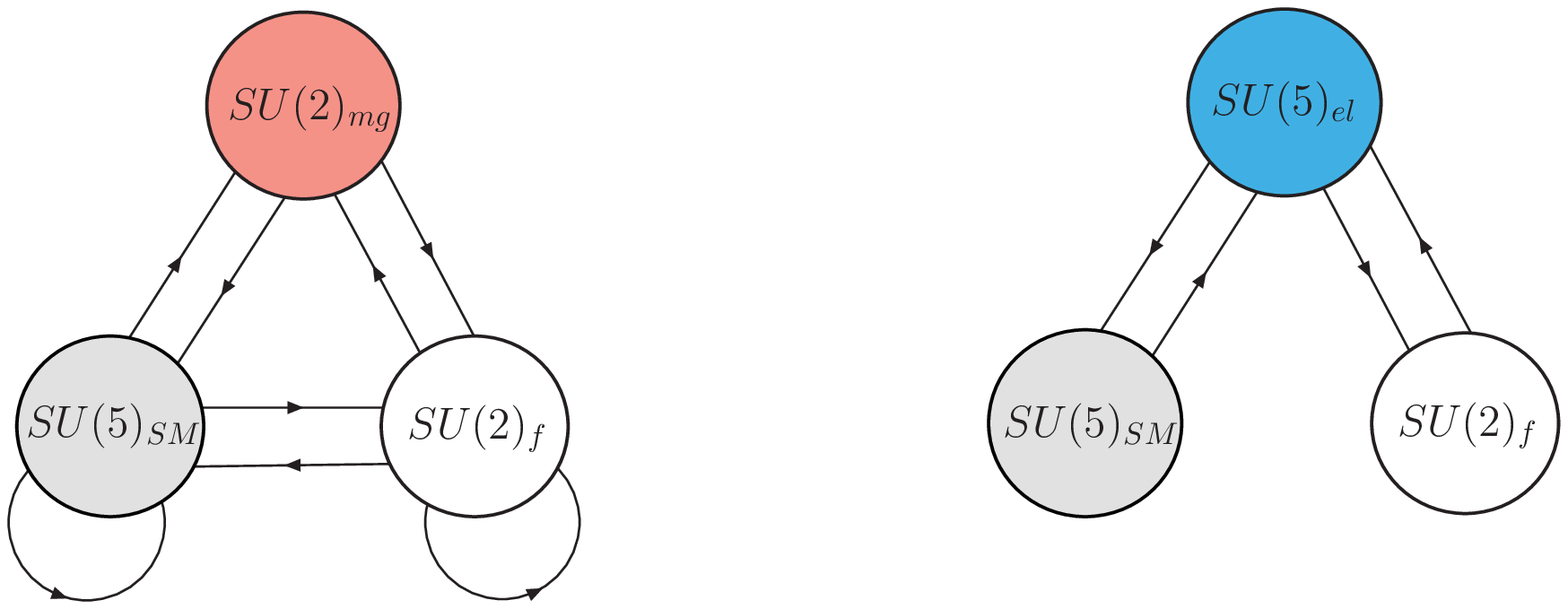}
\end{centering}
\caption{\it Quiver diagrams for the direct gauge mediation model of 
ref.\cite{us}, showing the ISS hidden sector in the magnetic (left) and 
electric (right) regimes. The coloured nodes represent the gauge group
of the hidden ISS sector, and the SSM parent gauge group. The blank nodes are 
left over ungauged flavour groups. The links between $SU(5)_{SM}$ and 
global $SU(2)_f$ are the composite mesons of the magnetic theory 
corresponding to independent directions in the electric theory moduli space.
The two regimes are matched at the scale $\sim \Lambda_{ISS}$. The contribution of the 
messengers to the $\beta$-functions of the SSM is 
$\Delta {b}_{SM}=-5-2-2 = -9$ in the magnetic regime 
but only 
$\Delta {b}_{SM}=-5$ in the electric regime. 
}
\end{figure}

The fields
and charges of the magnetic formulation are shown in 
Table~\ref{fieldstable},
\begin{table}
\begin{center}
\begin{tabular}{|c|c|c|c|c|}
\hline
&{\small $SU(2)_{mag}$} &
{\small $SU(2)_f$}&
$SU(5)_f$&
{\small $U(1)_{R}$}\tabularnewline
\hline
\hline
$\Phi_{ij}\equiv\left(\begin{array}{cc}
Y & Z\\
\tilde{Z} & X\end{array}\right)$&
{\bf 1}&
$\left(\begin{array}{cc}
Adj +{\bf 1} & \bar\square\\
\square & {\bf 1}\end{array}\right)$&
$\left(\begin{array}{cc}
{\bf 1} & \square\\
\bar{\square} & Adj+{\bf 1}\end{array}\right)$&
2
\tabularnewline
\hline
{\small $\varphi\equiv\left(\begin{array}{c}
\phi\\
\rho\end{array}\right)$}&
$\square$&
$\left(\begin{array}{c}
\bar{\square}\\ {\bf 1} \end{array}\right)$&
$\left(\begin{array}{c}
{\bf 1}\\
\bar{\square}\end{array}\right)$&
$1$
\tabularnewline
\hline
{\small $\tilde{\varphi}\equiv\left(\begin{array}{c}
\tilde{\phi}\\
\tilde{\rho}\end{array}\right)$}&
$\bar\square$&
$\left(\begin{array}{c}
\square\\ {\bf 1} \end{array}\right)$&
$\left(\begin{array}{c}
{\bf 1}\\
\square\end{array}\right)$&
$-1$\tabularnewline
\hline
\end{tabular}
\end{center}
\caption{\it Matter fields of the magnetic theory in \eqref{Wbardef}
and their decomposition under the gauge $SU(2)_{mag}$, the flavour $SU(2)_f \times SU(5)_f$ symmetry,
and their charges under the $R$-symmetry.
\label{fieldstable}}
\end{table}
and the superpotential is given by 
\begin{equation}
\label{Wbardef}
W=\Phi_{ij}\varphi_{i}.\tilde{\varphi_{j}}-\mu_{ij}^{2}\Phi_{ji}+m\varepsilon_{ab}\varepsilon_{rs}\varphi_{r}^{a}\varphi_{s}^{b}\,\,
\end{equation}
where $i,j=1...7$ are flavour indices, $r,s=1,2$ run over the first
two flavours only, and $a,b$ are $SU(2)_{mg}$ 
indices (we set the Yukawa coupling to unity for simplicity). 
By a gauge and flavour rotation, the matrix
$\mu_{ij}^{2}$ can be brought to a diagonal form 
\begin{equation}
\mu_{ij}^{2}=\left(\begin{array}{cc}
\mu_2^{2}\mathbf{I}_{2} & 0\\
0 & {\mu}_5^{2}\mathbf{I}_{5}\end{array}\right).
\end{equation}
We will assume that $\mu_2^{2}> {\mu}_5^{2}$. As explained in 
\cite{iss}, the ``rank-condition'' leads to metastable 
SUSY breaking such that $F_{X}= {\mu}_2^2$. The parameters $\mu_2^{2}$,
${\mu}_5^{2}$ 
and $m$ have an interpretation in terms of the electric
theory: $\mu_2^{2}\sim\Lambda_{ISS} m_{Q_2}$ and ${\mu}_5^{2}\sim\Lambda_{ISS} {m}_{Q_5}$
come from the electric quark masses $m_{Q_2}$, ${m}_{Q_5}$, where
$\Lambda_{ISS}$ is the dynamical scale of the ISS sector. 
The last term in \eqref{Wbardef} is the baryon deformation of the ISS model. 
As explained in \cite{us,us2} it is needed to trigger spontaneous breaking of the $R$-symmetry
 -- by generating $\langle X \rangle \neq 0$ as well as $\langle Y\rangle$, $\langle \varphi\rangle$, $\langle \tilde{\varphi}\rangle$ 
  --  required 
for non-zero gaugino masses in SSM.
This baryon operator can
be identified with a corresponding operator in the electric theory.

In this model there is no separate mediating sector, but the $SU(7)_f$
flavour symmetry is explicitly broken to 
$SU(2)_f\times SU(5)_f$.  The 
$SU(5)_f$ subgroup is gauged and associated with the parent $SU(5)$
of the Standard Model. The matter fields charged under this $SU(5)$ play the role of direct messengers;
these are the
magnetic quarks $\rho,\tilde{\rho}$ together with the meson components $X$, $Z$ and $\tilde{Z}$.

Now, as frequently occurs in direct mediation there is a Landau pole in
the SSM as well as the ISS sector, because of the large number of
additional flavours. Indeed an estimate was made in ref.\cite{us}
of where this occurs. In the magnetic theory the $SU(2)_{mg}$ magnetic
quarks $\rho$, $\tilde{\rho}$ contribute $-2$ to the 
$\beta$-function, and the $7\times7$ magnetic mesons $\Phi$ contribute $-2-5=-7$
(the $-2$ coming from the off-diagonal entries, $Z$, $\tilde{Z}$ and 
the $-5$ from the adjoint, $X$). Thus one can estimate 
\begin{equation}
b_{A}=b_{A}^{(MSSM)}-9
\end{equation}
and hence 
\begin{equation}
\alpha_{A}^{-1}=(\alpha_{A}^{-1})^{(MSSM)}-{9}\log(Q/\mu_2)
\end{equation}
where $\mu_2$ is the effective messenger scale, and where here and throughout 
we will be using for convenience the convention that 
\[
\ialpha \equiv \frac{8\pi^2}{g^2}
\]
rather than the more usual $4\pi/g^2$. 
To avoid an SSM Landau pole
before unification one requires\begin{equation}
(\alpha_{GUT}^{-1})^{(MSSM)}\gtrsim {9}\log(M_{GUT}/\mu_2)\end{equation}
or $\mu_2\gtrsim10^{9}$GeV, which is orders of magnitude above what
one wants for normal gauge mediation. (Indeed such a high value is 
close to the gravity mediation scale.)

However this estimate takes no account of the change of slope
in the electric ISS formulation, which is the appropriate description of the Hidden sector above the scale $\Lambda_{ISS}$. 
Indeed the lower limit on the latter
is only of order $10^{6}$GeV \cite{us2}, 
and above the scale $\Lambda_{ISS}$ the contribution to the SSM 
$\beta$-functions comes from the $N_c=5$ ``flavours'' of electric
quarks and antiquarks, and is just $-5$. (The mesons are composite objects in the electric dual and do not contribute
to the SSM $\beta$-functions as independent degrees of freedom.)
Taking this change of slope
into account, the gauge couplings are therefore 
\begin{equation}
\alpha_{A}^{-1}=(\alpha_{A}^{-1})^{(MSSM)}-{9}\log(\Lambda_{ISS}/\mu_2)-{5}\log(Q/\Lambda_{ISS})\, .
\end{equation}
A Landau pole appears if 
\begin{eqnarray}
(\alpha_{GUT}^{-1})^{(MSSM)} & \lesssim & {9}\log(\Lambda_{ISS}/\mu_2)+
{5}\log(M_{GUT}/\Lambda_{ISS})\nonumber \\
 & = & {4}\log(\Lambda_{ISS}/\mu_2)+{5}\log(M_{GUT}/\mu_2).
 \end{eqnarray}
Clearly minimizing $\Lambda_{ISS}/\mu_2$ ameliorates the Landau pole,
so assuming that $\Lambda_{ISS}\sim 10^{1-3}\mu_2$ we require $\frac{5}{2\pi}\log(M_{GUT}/\mu_2)\lesssim 20$
to avoid Landau poles or \[
\mu_2\geq4\times 10^{5}\,\mbox{GeV.}\]
 This requirement is easily met by the phenomenological models of
ref.\cite{us2}. 

Thus slopes can change upon Seiberg dualizing, and in particular there
can be a \emph{reduction} of the effective number of messengers 
in a model of direct gauge mediation, that delays the onset of Landau poles
to beyond the GUT scale. This is a very simple example of how
duality and unification can be interrelated. The main point of interest
is that rather than the familiar case whereby some degrees of freedom
are ``integrated in'' at higher energy scales, there is instead
a reduction in the effective degrees of freedom.
 
One can generalize the discussion to arbitrary numbers of flavours and 
colours. The general
condition for avoiding a Landau pole in this scenario is
\begin{eqnarray}
(\alpha_{GUT}^{-1})^{(MSSM)} & \gtrsim & (b-\bar{b})\log\left(
\Lambda_{ISS}/\mu_2\right)+
(b^{(MSSM)}-b)\log\left( M_{GUT}/\mu_2 \right)
\end{eqnarray}
where $\bar{b}$($b$) are the $\beta$-functions of the 
magnetic(electric) theories. For example, generalizing the 
model above so that the SSM is embedded in $SU(N_c)$ and there are
$N_f$ flavours in the supersymmetry breaking ISS sector gives the condition 
\begin{eqnarray}
(\alpha_{GUT}^{-1})^{(MSSM)} & \gtrsim & 2(N_f-N_c)\log \left( \Lambda_{ISS}/\mu_2 \right)+
N_c\log\left(M_{GUT}/\mu_2\right)\, .
\end{eqnarray}
Note that the matching of the electric and magnetic theories in the 
running of the one loop gauge coupling is well understood. Indeed the 
electric and magnetic hidden sector gauge 
theories have dynamical scales related by a matching relation such as
\be
\bar{\Lambda}^{\bar{b}} \Lambda^{b} = (-1)^{N_f-N_c} {\mu}^{N_f} 
\, , \label{matching0}
\ee
where $\mu$ is an undetermined scale (not to be confused with $\mu_{ij}$ in \eqref{Wbardef}) relating the composite  
mesons of the electric theory $M_{ij}=Q_i. \tilde{Q}_j$ 
to the elementary meson in the magnetic theory $M_{ij}=\mu \Phi_{ij} $.
Since $\mu $ is unknown, we may choose it 
so that 
$\Lambda < \bar{\Lambda}$ and there is a perturbative overlap of the 
two theories.

\section{An example of visible sector dual-unification}

We now turn to the complementary class of scenarios where the dynamics of the Hidden sector plays little or no
role in the Landau pole problem of the Visible sector.
Consider the possibility that the SSM is itself
a magnetic dual theory which becomes strongly coupled and develops Landau poles above the messenger scale. 
As we have said, this is natural in many scenarios of gauge mediation, and 
if the mediating fields appear in complete $SU(5)$ multiplets the SSM unification 
still occurs but at unphysical values of the gauge couplings. 
If this is the case, is it possible that this "fake" gauge 
unification is simply a manifestation of a real unification taking place
in an electric dual theory?
In fact there are examples in the literature of electric/magnetic dual GUT theories, which 
we can examine to answer this question (in the positive of course). As our prototypical example we 
will look at the model of Kutasov, Schwimmer and Seiberg (KSS)
\cite{KS,KS2} (see refs.\cite{poppitz,poppitz2,klein2,klein} for related work). 
In the following two subsections we briefly review its details, and then  
discuss dual-unification in subsection 3.3 and present explicit examples with magnetic $SU(5)$ in 3.4. 

\subsection{The electric theory}

The microscopic theory is an $SU(N_{c})$ gauge theory with $N_{f}$ flavours of quarks $Q$
and anti-quarks $\tilde{Q}$, and an adjoint field of the $SU(N_{c})$, $X$. 
The superpotential defining the model is
\be
W= \sum_{i=0}^k \frac{s_i}{k+1-i} \Tr X^{k+1-i} + \lambda \Tr X
\ee
where $s_0, \ldots, s_k$ are constants and $k$ is a fixed integer. 
The constant $\lambda$ is the Lagrange multiplier
which ensures tracelessness of $X$. The leading term in $W$ (i.e. the term with the highest power of $X$) is
$\frac{s_0}{k+1} \Tr X^{k+1}$, and the subleading terms with $i>0$ are often thought of as deformations. 
The parent global symmetry of the theory when only the leading $s_0$ term is present is 
\[
SU(N_{f})\times SU(N_{f})\times U(1)_{B}\times U(1)_{R}
\]
with the charges shown in Table~\ref{fieldstable2}.
\begin{table}
\begin{center}
\begin{tabular}{|c||c|c|c|c|}
\hline 
$ $ & $SU(N_{f})$ & $SU(N_{f})$ & $U(1)_{B}$ & $U(1)_{R}$\tabularnewline
\hline
\hline 
$Q$ & $N_{f}$ & 1 & 1 & $1-\frac{2}{k+1}\frac{N_{c}}{N_{f}}$\tabularnewline
\hline 
$\tilde{Q}$ & 1 & $\bar{N_{f}}$ & -1 & $1-\frac{2}{k+1}\frac{N_{c}}{N_{f}}$\tabularnewline
\hline 
$X$ & 1 & 1 & 0 & $\frac{2}{k+1}$\tabularnewline
\hline
\end{tabular}
\par\end{center}
\caption{Matter fields of the microscopic KSS theory and their global charges.
\label{fieldstable2}}
\end{table}
When there are non-zero $s_{i>0}$ the $R$-symmetry is 
completely broken. These deformations are responsible for generating the VEV for $X$ and spontaneously 
breaking the $SU(N_c)$ gauge symmetry as we shall see. 
KSS also use an
equivalent form of this superpotential,
\be
W= \sum_{i=0}^k \frac{t_i}{k+1-i} \Tr X^{k+1-i} + \lambda' \Tr X
\ee 
where $X$ is shifted by $\frac{s_1}{s_0 k} {\bf 1}$ 
and the constants $t_i$ and $\lambda'$ are chosen 
in terms of $s_i$ and $\lambda$ so that the coefficient of the first subleading term $\Tr X^{k}$ is zero (with $s_0=t_0$). 
This form is useful for ensuring that traceless adjoints in the magnetic theory are consistent with
the vacuum structure.

The $F_{X}$-term
equation for non-zero $s_{i}$'s can easily be solved by diagonalizing
the $X$ using $SU(N_{c})$ rotations and dictates the vacuum structure; 
the equation for a single entry $x$ on the diagonal is 
\be
W'=0\equiv\sum_{i=0}^{k-1}s_{i}x^{k-i}+\lambda.
\ee
This is a $k^{\mbox{th}}$ order polynomial so there are $k$ roots; hence 
\be
\langle X\rangle=\left(\begin{array}{cccc}
x_{1}\mathbf{I}_{r_{1}}\\
 & x_{2}\mathbf{I}_{r_{2}}\\
 &  & ...\\
 & & & x_{k}\mathbf{I}_{r_{k}}\end{array}\right)
 \ee
where 
\be
\sum_{i=1}^k r_i=N_c\, .
\ee
The original microscopic gauge group is broken (Higgsed) down to \be
SU(N_{c})\rightarrow SU(r_{1})\times SU(r_{2})\ldots SU(r_{k})\times U(1)^{k-1}.
\label{elHiggs}
\ee
with the values of the roots $x_i$ being fixed in terms of 
the $s_i$'s (these are not flat directions). 
Around each such vacuum the adjoint field $X$ is massive and can 
be integrated out. The resulting theory
at energy scales below $\langle X \rangle$ (or  more precisely, 
below the scales set by the differences $x_i-x_j$)
is the product of $k$ ordinary SQCD theories times Abelian SQED-like factors $U(1)^{k-1}$ (which are essentially gauged baryon numbers).

The microscopic electric $SU(N_c)$ will here play the role of 
the Grand Unified Theory; the unification above
the $M_{GUT} \sim x_i-x_j$ scale(s) is by default, since 
we started from the single $SU(N_c)$ gauge group.
For simplicity we will always assume that the original 
$SU(N_c)$ theory is well-defined in the UV, i.e. is asymptotically free.
Its one-loop $\beta$-function coefficient 
\be 
b_0 = 3N_c - N_f - N_c = 2N_c- N_f > 0
\ee
must therefore be positive. 
(Here $-N_c$ comes from the adjoint field $X$.) The dynamical 
transmutation scale of this theory we will take to be $\Lambda$.

\subsection{The magnetic theory}

Now let us go to the macroscopic theory, starting with the dual of the 
\emph{unbroken} $SU(N_c)$ microscopic theory. To this end
we can (almost) switch off the VEVs, i.e. we assume that the `GUT'-scale 
set by $\langle X \rangle$, is very small (and in particular, 
is much below $\Lambda$). This can always be achieved by
appropriately dialing down the constants $s_{i>0}$ in the superpotential. 
The magnetic dual theory has the gauge group
$SU(\bar{N}_c)$ with $N_f$ flavours of magnetic quarks and antiquarks 
$q$ and $\tilde{q}$, a set of $k$ mesons $M_j$,
and an adjoint field $Y$ (as described in refs.\cite{KS,KS2}), where
\be
\bar{N}_c = k N_f - N_c\, .
\ee
The $k$ mesons in the electric theory
are composites of the electric quarks and the adjoint field $X$, 
\begin{equation}
M_{j}=\tilde{Q}X^{j-1}Q\,\,;\,\,\, j=1\ldots k.
\end{equation}
In the magnetic theory these mesons are included as fundamental (non-composite) degrees of freedom.
The $j=1$ object is the usual meson, and in fact the $k=1$ model
is just the usual Seiberg's magnetic SQCD. 

The one-loop $\beta$-function of the magnetic theory is
\be
\bar{b}_0 = 2\bar{N}_c- N_f \, ,
\ee
and the corresponding dynamical scale $\bar{\Lambda}$ 
is related to $\Lambda$ of the electric theory
via the matching relation \cite{KS2}
\be
\bar{\Lambda}^{\bar{b}_0} \Lambda^{b_0} = \left(\frac{\mu}{s_0}\right)^{2N_f} 
\, . \label{matching1}
\ee
Here $\mu$ is the scale required for relating the operators of the theory in the UV to the IR -- 
recall that scaling dimensions
of various operators are not the same in the UV and the IR. This scale is undetermined (beyond the equation \eqref{matching1})
by arguments based on duality and holomorphicity, its value depending on the nonholomorphic K\"ahler potential over which 
we have little control. However, as soon as $\mu$ is known, $\bar{\Lambda}$ is determined through eq.\eqref{matching1}.
This equation is uniquely fixed by the transformation properties under the
global symmetries of the undeformed ($s_{i>0}=0$) theory which are 
shown in  
Table~\ref{fieldstable3}.
\begin{table}
\begin{center}
\begin{tabular}{|c||c|c|c|c|}
\hline 
$ $ & $SU(N_{f})$ & $SU(N_{f})$ & $U(1)_{B}$ & $U(1)_{R}$\tabularnewline
\hline
\hline 
$q$ & $N_{f}$ & 1 & $\frac{N_{c}}{\bar{N_{c}}}$ & $1-\frac{2}{k+1}\frac{\bar{N_{c}}}{N_{f}}$\tabularnewline
\hline 
$\tilde{q}$ & 1 & $\bar{N_{f}}$ & $-\frac{N_{c}}{\bar{N_{c}}}$ & $1-\frac{2}{k+1}\frac{\bar{N_{c}}}{N_{f}}$\tabularnewline
\hline 
$Y$ & 1 & 1 & 0 & $\frac{2}{k+1}$\tabularnewline
\hline
$M_{j}$ & $N_{f}$ & $\bar{N_{f}}$ & 0 & $2-\frac{4}{k+1}\frac{N_{c}}{N_{f}}+\frac{2}{k+1}(j-1)$\tabularnewline
\hline
\end{tabular}
\par\end{center}
\caption{Matter fields of the magnetic KSS theory and their global charges.
\label{fieldstable3}}
\end{table}

\noindent The full superpotential in the {\em deformed} magnetic theory is of the form
\begin{equation}
W_{mag}=\sum_{i=0}^{k-1}\frac{-{t}_{i}}{k+1-i}Tr(Y^{k+1-i})+\frac{1}{\mu^{2}}\sum_{l=0}^{k-1}t_{l}
\sum_{j=1}^{k-l}M_{j}\tilde{q}Y^{k-j-l}q + \mbox{const}
\label{eq:wmag}\end{equation}
where the Lagrange
multiplier term has already been determined.
As soon as the deformation is turned on, and the electric $SU(N_c)$ theory 
is Higgsed in accordance with \eqref{elHiggs}, 
the magnetic $SU(\bar{N}_c)$ theory is broken as well by the magnetic 
adjoints acquiring VEVs of the form \cite{KS,KS2}
\be
\langle Y \rangle = \left(\begin{array}{cccc}
y_1 {\mathbf I}_{\bar{r}_1}\\
& y_2 {\mathbf I}_{\bar{r}_2}\\
&  & ...\\
 & & &  y_k {\mathbf I}_{\bar{r}_k}\end{array}\right)
\label{Yvevs}
\ee
with 
\be 
y_i-y_j = x_i - x_j \, .
\label{yxvevs}
\ee
Thus for the purposes of this paper, we have two approximate 
scales defining the theory: the scale of symmetry 
breaking $M_{GUT}\sim x_i-x_j$, with $t_{i}\sim M_{GUT}^{i+2-k}$,
and the scale $\mu$. The corresponding symmetry breaking is
\be
SU(\bar{N}_c) \to SU(\bar{r}_1) \times \ldots \times SU(\bar{r}_k) \times U(1)^{k-1}
\label{mgHiggs}
\ee
where 
\be
\bar{r}_i= N_f - r_i \ , \qquad \sum_{i=1}^k \bar{r}_i = \bar{N}_c.
\ee
This magnetic theory breakdown is very similar to the electric theory breaking, and is ensured by the
form of the magnetic superpotential \eqref{eq:wmag}. In particular the coefficients 
of eq.\eqref{eq:wmag} are 
(up to a common factor) determined by
checking that the vacuum structure matches in the manner described 
by eq.(\ref{yxvevs}) and 
that, for example, critical points coincide in both theories 
simultaneously. (The matching is greatly simplified by the choice of 
$t_i$ coordinates rather than the original $s_i$ coordinates.)

For the most part, the $SU(r_i)\leftrightarrow SU(\bar{r}_i)$ duality in the 
broken theory is exactly the normal Seiberg duality of SQCD. Thus in 
this particular class of models the Higgsing has 
broken the whole theory into a decoupled product of SQCD Seiberg duals. 
However, one remarkable exception is the unbroken electric
theory mapping onto a broken magnetic theory. If we choose $k=2$ and 
$r_{1}=N_{c}$,
then the electric theory is clearly unbroken but the magnetic theory
is broken as \begin{equation}
SU(\bar{N}_{c})\rightarrow SU(N_{f}-N_{c})\times SU(N_{f})\times U(1).\end{equation}

The final ingredient we will need from the models of Ref.\cite{KS,KS2}, 
is the matching conditions
for the dynamical scales of the constituent $SU(r_i)$ and the $SU(\bar{r}_i)$ 
factors of the electric and the magnetic theories; these are given by \cite{KS2}
\be
\bar{\Lambda}_i^{\bar{b}_i} \Lambda_i^{b_i} = (-1)^{N_f-r_i} \mu_i^{\bar{b}_i+b_i},
\label{matching2}
\ee
where the scale $\mu_i$ is determined to be
\be
\mu_i = \frac{\mu^2}{t_0}\frac{1}{\prod_{i \neq j} (x_i-x_j)}
\label{muscale}
\ee
and the $\beta$-function coefficients of the electric and magnetic SQCD factors are\footnote{Note 
that here the adjoint fields have already been integrated out.}
\be
b_i = 3 r_i - N_f \ , \qquad \bar{b}_i = 3 \bar{r}_i - N_f.
\ee
It will be convenient to write \eqref{matching2} as
\be
\label{matching3}
b_i t_{\Lambda_i} + \bar{b}_i t_{\bar{\Lambda}_i} = (b_i +\bar{b}_i) t_{\mu_i}
\ee
where $t_E \equiv \log E$. Here we have ignored a possible phase factor which only affects the $\theta_{YM}$ parameter.
It follows from \eqref{matching3} and the one-loop definition of the dynamical transmutation scale, 
\be 
\ialpha(t) = b(t-t_\Lambda)\, ,
\label{lambdef}
\ee
that the physical meaning of $\mu_i$ is that it is the scale where 
the one-loop electric and magnetic couplings are equal and opposite, 
$\ialpha_i(\mu_i) = - \bar{\alpha}^{-1}_i(\mu_i)$. 
A negative coupling in this context 
implies that the corresponding theory is strongly coupled and that the perturbative description of the theory is invalid.

The matching conditions \eqref{matching2} and \eqref{muscale} can be 
derived in two ways, as follows. Either one can consult the superpotential 
\eqref{eq:wmag}, integrate out the massive adjoints fields and use the 
well known SQCD matching condition for each $SU(r_i)$ factor. 
The scale $\mu_{i}$ is then determined in the usual way as the coupling to the
meson in the magnetic superpotential term\begin{equation}
W_{mag}^{(i)}=\frac{Q\tilde{Q}}{\mu_{i}}q\tilde{q}.\end{equation}
On integrating out the adjoints, eqs.\eqref{matching2} and \eqref{muscale} follow straightforwardly 
from eq.\eqref{eq:wmag}.
Alternatively, one can match the dynamical transmutation scales of the 
broken and unbroken theories on the magnetic and electric sides, and relate 
them via eq.\eqref{matching1}. For example on the electric side at 
a scale $E_i$ at
which some degrees of freedom are integrated out, using the usual
Wilsonian expression for the scale dependence of the gauge couplings (or equivalently 
eq.\eqref{lambdef}), one has
\begin{equation}
e^{-\frac{8\pi^{2}}{g^{2}(E_i)}}=\left(\frac{\Lambda_i}{E_i}\right)^{b_i}
= \left(\frac{\Lambda'_i}{E_i}\right)^{b_i'} ,\label{eq:wilson}\end{equation}
where $\Lambda_i'$ is the new dynamical scale in the theory without those particular degrees of freedom. In this way we can work our way back to the unbroken theory 
and relate its $\Lambda$ to that of the theory with all the heavy degrees of freedom integrated out in both the electric and magnetic theories:
\begin{eqnarray} 
\Lambda_i^{b_i} &= & \Lambda^{b_0} t_0^{r_i} \prod _{j\neq i} (x_i-x_j)^{r_i-2r_j}\nonumber \\
\bar{\Lambda}_i^{\bar{b}_i} &= & \bar{\Lambda}^{\bar{b}_0} t_0^{\bar{r}_i} \prod _{j\neq i} (y_i-y_j)^{\bar{r}_i-2\bar{r}_j}\, .
\label{lambdas}
\end{eqnarray}
Multiplying these two and using eqs.\eqref{matching1} and \eqref{yxvevs}
one arrives at 
eq.\eqref{matching2} \cite{KS2}. The fact that the two answers agree is 
a nontrivial check of the superpotential. 

We should add a clarifying remark about the veracity of this procedure
in the present context. We will be interested in magnetic 
theories that are IR free, matching onto  electric theories that are asymptotically free. But 
the states we are integrating out have masses $x_i-x_j\sim M_{GUT}$, and so
the two equations \eqref{lambdas} 
are derived for different regimes of $M_{GUT}$ (either $M_{GUT}>\Lambda$ for the electric theory, or  $M_{GUT}<\bar{\Lambda}$ for the magnetic theory). What makes it valid to multiply them to get 
eq.\eqref{matching2} is analyticity. The dynamical transmutation scale 
determines (the real part of) the gauge kinetic function, and the parameter $\mu$ appears in the 
superpotential. Therefore, since the VEVs of the fields are functions of the coordinates, $t_i$, 
the derivation of eq.\eqref{matching2} is a form of analytic continuation in the $t_i$.

\subsection{Dual-unification}
\label{sect:dual-unification}

Now, let us see that dual-unification actually 
happens for the KSS model \cite{KS,KS2}. 
First the unification
in the electric theory means via eq.\eqref{lambdef} 
that
 \begin{equation}
\ialpha_{i}(t)=\ialpha_{GUT}+b_{i}(t-t_{GUT})\label{eq:alpha-elec}\end{equation}
where the $i$-suffix indicates the particular group factor\footnote{This can be taken as the definition of simple unification (at one-loop), meaning that
all electric couplings $\alpha_{i}(t_{GUT})$ are the same (with no threshold effects).}. Our 
aim in this subsection is to show
that by matching the electric to the magnetic theories in the correct
way, the magnetic theory couplings can be written as\begin{equation}
\bar{\alpha}^{-1}_{i}(t)=\bar{\alpha}^{-1}_{GUT}+\bar{b}_{i}(t-t_{GUT}),
\label{eq:alpha-mag}
\end{equation}
where $\bar{\alpha}_{GUT}$ is a new effective unification value
that will be negative. Thus the magnetic theory \emph{looks }as if
it is unifying at negative values of $\bar{\alpha}_{i}$. 

The magnetic couplings are \begin{eqnarray}
\bar{\alpha}^{-1}_{i}(t) & = & 
\bar{b}_{i}(t-t_{\bar{\Lambda}_{i}})\nonumber \\
 & = & \bar{b}_{i}(t_{GUT}-t_{\bar{\Lambda}_{i}})+\bar{b}_{i}(t-t_{GUT})\label{eq:alpha-mag2}\end{eqnarray}
so we must determine the first piece and ensure that it is
independent of $i$ (i.e. that unification occurs in the magnetic theory).
It is easy to see that the matching condition \eqref{matching2} ensures this; writing it as 
\begin{equation}
b_{i}t_{\Lambda_{i}}+\bar{b}_{i}t_{\bar{\Lambda}_{i}}=(b_{i}+\bar{b}_{i})t_{\mu_{i}},
\label{eq:matching2}\end{equation}
and using 
\eqref{lambdef} we can recast the first piece of 
eq.\eqref{eq:alpha-mag2} as follows:
\begin{eqnarray}
\bar{b}_{i}(t_{GUT}-t_{\bar{\Lambda}_{i}}) & = & \bar{b}_{i}t_{GUT}-(b_{i}+\bar{b}_{i})t_{\mu_{i}}+b_{i}t_{\Lambda_{i}}\nonumber \\
 & = & \bar{b}_{i}t_{GUT}-(b_{i}+\bar{b}_{i})t_{\mu_{i}}-\ialpha_{GUT}+b_{i}t_{GUT}\nonumber \\
 & = & (b_{i}+\bar{b}_{i})(t_{GUT}-t_{\mu_{i}})-\ialpha_{GUT}\, . \end{eqnarray}
Finally since \begin{equation}
(b_{i}+\bar{b}_{i})=N_{f}\end{equation}
we can write \begin{equation}
\bar{\alpha}^{-1}_{i}(t)=\bar{\alpha}^{-1}_{GUT}+
\bar{b}_{i}(t-t_{GUT}),\end{equation}
as required, where \begin{equation}
\bar{\alpha}^{-1}_{GUT}=N_{f}(t_{GUT}-t_{\hat{\mu}})-\alpha^{-1}_{GUT},\label{eq:alpha-gut}\end{equation}
and where we have defined a new scale which should be common to all 
the $SU(r_i)$ subgroups:
\begin{equation}
\hat{\mu}\sim\mu_i = \frac{\mu^{2}}{t_{0}}\frac{1}{\prod_{j\neq i}(x_{i}-x_{j})}\, ,
\quad \forall \, i \, .\end{equation}
Note that \eqref{eq:alpha-gut} is simply the statement that $\hat{\mu}$ is the scale where
\be
\bar{\alpha}(\hat{\mu})^{-1}=-\alpha(\hat{\mu})^{-1}\,  .
\ee
Thus the dual-unification is only realized if the $\mu_{i}$ are all of the same order; but this 
is also required for simple unification in the electric theory since the $\mu_i$ are
derived from the masses of the heavy states that we integrate out to find the matching condition. If these states are not degenerate then simple unification does not happen 
in either the electric or magnetic theories (although of course one can still have unification with a number of thresholds).  

The point is that as long as all the $x_{i}$ and hence $\mu_{i}$
are of the same order of magnitude, 
the variation in $\bar{\alpha}_{GUT}$
is a threshold effect. 
For the $SU(\bar{r}_i)$ factors, the fact that there is unification doesn't
depend on the scale at which the adjoint fields are integrated out
(this is given by $t_{0}$), although of course $\bar{\alpha}_{GUT}$
does. This is because the states are still integrated out in complete $SU(N_c)$ 
multiplets for any value of $t_0$. 
In order to have a complete unification, it seems natural to suppose 
that all the parameters of the deformed 
superpotential in the electric theory are determined by a 
single scale, $M_{GUT}$. Under this
assumption we can estimate \begin{eqnarray}
\hat{\mu} & \approx & \frac{\mu^{2}}{M_{GUT}}\end{eqnarray}
 and hence \begin{equation}
\bar{\alpha}^{-1}_{GUT}=2N_{f}(t_{GUT}-t_{\mu})-\ialpha_{GUT}.\label{eq:alpha-bar2}\end{equation}

So far we have established that \eqref{eq:alpha-mag} holds for the non-abelian factors of the 
magnetic theory, with the universal value of $\bar{\alpha}_{GUT}$ being given by 
\eqref{eq:alpha-bar2}. In a moment we shall show that the $U(1)$ factors of the magnetic theory 
unify with the non-abelian factors at the same scale. Before we do this however, let us 
check that eq. (\ref{eq:alpha-bar2}) is consistent with the
relation between the dynamical scales of the two unbroken theories. Indeed in
deriving it we assumed only that $\mu_{i}<\Lambda_{i},\,\bar{\Lambda}_{i}$,
and did not require any information about the relative magnitude of
$M_{GUT}.$ Therefore one can go to the limit where the
GUT symmetry breaking is turned off.
Indeed eq.\eqref{matching1} relating the 
dynamical scales of the unbroken theories gives \begin{eqnarray}
(2N_{c}-N_{f})t_{\Lambda}+(2(kN_{f}-N_{c})-N_{f})t_{\bar{\Lambda}} & = & 2N_{f}(t_{\mu}-t_{s_{0}})\nonumber \\
 & \approx & 2N_{f}(t_{\mu}-(2-k)t_{GUT}),\label{eq:ks2 relation}\end{eqnarray}
 where we have used $s_0\sim M_{GUT}^{2-k}$. 
Using $\ialpha_{GUT}=\ialpha(t_{GUT})=(2N_{c}-N_{f})(t_{GUT}-t_{\Lambda})$, this can be written
\begin{eqnarray}
(2N_{c}-N_{f})t_{GUT}-\ialpha_{GUT}+\,\,\,\,\,\,\,\,\,\,\,\,\nonumber \\
\,\,\,\,\,(2(kN_{f}-N_{c})-N_{f})t_{GUT}-\bar{\alpha}^{-1}_{GUT} & = & 2N_{f}(t_{\mu}-t_{s_{0}})\\
 & \approx & 2N_{f}(t_{\mu}-(2-k)t_{GUT}).\nonumber \\
\Longrightarrow\ialpha_{GUT}+\bar{\alpha}^{-1}_{GUT} & = & 
2N_{f}(t_{GUT}-t_{\mu})\, ,\end{eqnarray}
which reproduces \eqref{eq:alpha-bar2} in the unbroken theory as well. In this sense the GUT symmetry breaking and the electric/magnetic
duality are permutable; the scale of the former may be continuously
dialed down until the physical phenomenon is best described by fundamental
excitations of the magnetic theory with unification occuring there instead; however unification is manifest in both descriptions. 

Finally let us turn to the abelian factors
and show that they also unify (in the GUT normalization) at the same scales as
the $SU(r_{i})$ factors. We see this as follows: the $\beta$-functions
of the $U(1)$'s in both the electric and magnetic theories are simply
$b_{U(1)}=\bar{b}_{U(1)}=-N_{f}$, so upon integrating out the heavy
($M_{GUT}$) degrees of freedom, the dynamical scales of the $U(1)$ factors
are related by eq.\eqref{eq:wilson} to those of the parent $SU(N_{c})$ (or
$SU(kN_{f}-N_{c})$) theory as \begin{eqnarray}
-N_{f}t_{\Lambda_{U(1)}} & = & (2N_{c}-N_{f})t_{\Lambda}-2N_{c}t_{GUT}\nonumber \\
-N_{f}t_{\bar{\Lambda}_{U(1)}} & = & (2(kN_{f}-N_{c})-N_{f})t_{\bar{\Lambda}}-2(kN_{f}-N_{c})t_{GUT}\,.\label{eq:landau}\end{eqnarray}
 Again using $\alpha^{-1}_{U(1)}(t_{GUT})=-N_{f}(t_{GUT}-t_{\Lambda_{U(1)}}$),
and its equivalent for the magnetic theory, we have \begin{eqnarray}
\label{eq:landau2}
-2N_{f}t_{GUT}-(\ialpha_{U(1)}(t_{GUT})+\bar{\alpha}^{-1}_{U(1)}(t_{GUT})) & = & (2N_{c}-N_{f})t_{\Lambda}+\nonumber \\
 &  & \,\,(2(kN_{f}-N_{c})-N_{f})t_{\bar{\Lambda}}-2kN_{f}t_{GUT}\,\nonumber \\
 & = & 2N_{f}(t_{\mu}-2t_{GUT})\end{eqnarray}
and hence \begin{equation}
\ialpha_{U(1)}(t_{GUT})+\bar{\alpha}^{-1}_{U(1)}(t_{GUT})=2N_{f}(t_{GUT}-t_{\mu})\end{equation}
as required for consistent unification. For the $U(1)$'s to unify with the $SU(r_i)$'s, a more careful analysis shows that 
we require $s_0\approx M_{GUT}^{2-k}$, because the electric and magnetic $U(1)$'s are not related 
by the duality in the same way as the $SU({r}_i)$ factors. This will be made explicit for a simple example in the 
next subsection.  

Note that, although they have the same slopes, the $U(1)$'s are in a sense 
dualized as well; they are inherited from the underlying GUT  $SU(N_c)$ and 
$SU(\bar{N}_c)$ theories which were dual to each other. 
This is an important point. One approach to finding dual-unification in the SSM 
would be to run the magnetic theory (which is what we have access to experimentally) until the first  
$SU(\bar{r}_i)$ factor becomes strongly coupled, perform a Seiberg duality on it and continue until 
all the $SU(\bar{r}_i)$ factors are dualized.
However complete unification involves the $U(1)$ factors as well. Here we showed that full knowledge 
of the GUT $SU(N_c)$ and $SU(\bar{N}_c)$ theories is required to see that all abelian and non-abelian factors unify correctly. 

A brief remark about the validity of Eq.(\ref{eq:landau}). This
equation is based on matching the broken and unbroken 
theories at the scale $M_{GUT}$
at which the degrees of freedom are integrated out, using eq.\eqref{eq:wilson}.
However we have already used
this relation for the dynamical scales of the $SU(r_{i})$ subfactors.
Is it valid to also use the same equation for the $U(1)$ factors
as well? The answer is yes, because we are matching different couplings
of a broken unified group. That is, we match the theories at the scale
$M_{GUT}$, where the integrating out of adjoint and vector degrees of freedom
effectively splits the $U(1)$ and the $SU(r_{i})$ running; however
these factors can be individually matched with the corresponding subgroups
of the unified theory via eq.\eqref{eq:wilson}.

\subsection{Simple examples with a magnetic $SU(5)$}

Now let us consider some specific examples, 
in order to make explicit these general features. We shall concentrate on 
$k=2$ since these are the first nontrivial cases, and also these are the most Standard-Model-like
of this particular class of theories. We shall choose the couplings in the superpotential
to be 
$s_0=1$, $s_1=m$ (note that $s_0$ can always be adjusted by renormalizing the 
adjoint fields, since we do not in general have 
canonical normalization) so that 
\be 
W=Tr(\frac{X^{3}}{3}+m\frac{X^{2}}{2}+\lambda X)\, .
\ee
The VEVs are 
\be
\langle X\rangle=\left(\begin{array}{cc}
x_{+}\mathbf{I}_{r_{+}}\\
 & x_{-}\mathbf{I}_{r_{-}}\\
 \end{array}\right)\, , 
 \qquad 
 r_++r_-=N_c\, .
\ee
The eigenvalues are 
\be
x_{\pm}=\frac{-m\pm\sqrt{m^{2}-4\lambda}}{2}\ee
and the condition $Tr(X)=0$ fixes 
\begin{eqnarray}
\lambda & = & -\frac{m^{2}}{4}\frac{r_+r_-}{(r_+-r_-)^2} \nonumber \\
x_{\pm} &= & \pm {m}\frac {r_\mp}{r_+-r_-}\, .
\end{eqnarray}
The masses of e.g. the fermions (note that supersymmetry is not being
broken here) are 
\be
W_{XX}=2X+m=\left\{ \begin{array}{cc}
m & \,\,\,\,\,;\,\, X\equiv X_{i\neq j}\\
\pm m\frac{N_{c}}{r_{+}-r_-} & \,\,;\,\, X\equiv X_{ii} \end{array}\right.\ee
Hence at scales below $m$ we can indeed 
just integrate out the adjoint fields
and are left with a product of $k$ usual SQCD-models. 
Of course this discussion is valid for both electric and magnetic 
theories with the obvious replacements.

For the example where $SU(5)$ 
splits into $SU(3) \times SU(2) \times U(1)$
we have $r_+=2$, $r_-=3$ and we get
\be
\lambda = -6 m^2\ , \qquad x_+ = 2m \ , \quad x_- = -3m
\ee
so that $\langle X \rangle = {\rm diag} (2m, 2m, 2m, -3m,-3m)$, as expected.
Since now there are only two roots 
$x_+$ and $x_-$, it follows that there is only one matching scale
for the case $k=2$ which makes it somewhat special.
Indeed, $\mu_i$ in \eqref{muscale} are now $\mu_+$ 
and $\mu_-$ such that (recalling that $t_0=s_0$ and that $s_0=1$ in this $k=2$ example)
\be
\mu_+ = -\mu_-= \frac{\mu^2}{(x_+-x_-)} = 
\frac{\mu^2}{m}\frac{(r_+-r_-)}{N_c},
\label{muscale2}
\ee
and the actual physical matching scale is the absolute value $|\mu_+|$.

It is easy to see that this unique scale makes 
the unification of the non-Abelian factors with the $U(1)$ factors exact in this case. 
In the electric theory this is true by construction, since $SU(N_c)$ was Higgsed down to $SU(r_+) \times SU(r_-)\times U(1)$
by the $\langle X \rangle$ VEV. In the magnetic theory, the fake unification (or dual-unification) follows from the
matching conditions, which for the non-Abelian factors were given by \eqref{matching2}. As we saw, the $U(1)$ factors are matched by integrating out degrees of freedom as in eq.\eqref{eq:landau}. A slightly more careful rendering 
of it (reinstating $t_0$) gives 
\be
\Lambda_{U(1)}^{-N_f} = {\Lambda^{b_0}} t_0^{N_c} |x_+-x_-|^{-2N_c} \ , \qquad
 \bar{\Lambda}_{U(1)}^{-N_f} = {\bar{\Lambda}}^{\bar{b}_0} t_0^{\bar{N}_c} |x_+-x_-|^{-2\bar{N}_c} \ ,
\ee
where $-N_f$ is the $\beta$-function of both $U(1)$ factors (in GUT normalisation) and we have ignored
the irrelevant (for our purposes) phase factor. This in turn gives 
\be
-N_f (t_{\Lambda_{U(1)}}+t_{\bar{\Lambda}_{U(1)}})= {2N_f} (t_\mu -2 t_{|x_+-x_-|})  \ , \qquad b_{U(1)} = - N_f = \bar{b}_{U(1)},
\ee
with the powers of $t_0$ cancelling. Comparison with eqs.\eqref{eq:landau} and \eqref{eq:landau2} 
shows that when $t_0 = 1$ the $U(1)$ coupling unifies with the other two couplings precisely
at $M_{GUT}=|x_+-x_-|$,  provided that the gauge couplings 
unify in the electric theory at that same scale. However when $t_0\neq 1$ the precise unification 
in the magnetic theory is spoiled by logarithmically small threshold effects, because of the explicit appearance
of $t_0$ in the matching condition for the $SU(r_i)$ factors in eq.\eqref{eq:ks2 relation}. This is to be expected 
since $t_0$ different from unity gives a
split mass spectrum. Simple unification is already lost in the electric theory.

Now, recall that the $\beta$-functions of the unbroken theories are \begin{eqnarray}
b_0 & = & 2N_{c}-N_{f}\nonumber \\
\bar{b}_0 & = & (2k-1)N_{f}-2N_{c}.\end{eqnarray}
Thus when the electric theory is asymptotically free (i.e.
$2N_{c}>N_{f}$) the magnetic theory need not be IR free (for example
if $N_{f}\lesssim2N_{c}$ then $\bar{b}_0\approx2(k-1)N_{f}$). The
same is true of the SQCD factors of the broken theory. Although all
of them are dualized some of them may be asymptotically free rather than IR free. 
As an example, consider a theory with $k=2$, $N_{f}=5$
and $SU(5)\rightarrow SU(3)\times SU(2)\times U(1)$ with $b_{SU(2)}=1$,
and $b_{SU(3)}=4$. This theory satisfies the condition for stable
vacua, $r_{i}\leq N_{f}$, but the magnetic theory is \begin{equation}
SU(5)\rightarrow SU(2)\times SU(3)\times U(1),\end{equation}
so it has the same gauge groups, number of flavours, and hence slopes.
In this case, all the SQCD factors of the broken theory are asymptotically free
 in both the electric and magnetic theories. Recall
that $\mu$ is the scale at which $\ialpha_{U(1)}(t_{\mu})=
-\bar{\alpha}^{-1}_{U(1)}(t_{\mu})$.
Since the slopes of both electric and magnetic $U(1)$'s are the same,
in order consistently to define the scale $\mu$ (i.e. with $\mu<M_{GUT}$)
we require $\bar{\alpha}_{GUT}<0$, which would mean that the couplings
of the magnetic theory are always unphysical. Equivalently, the unification
takes place in the magnetic phase. Such theories are irrelevant
to us.

 We will therefore focus on theories that have all
SQCD factors in the free magnetic range, in this case $\frac{3}{2}r_{i}>N_{f}\geq r_{i}+1\,\,\forall i$. 
We also require that the magnetic GUT theory is not asymptotically free 
while the electric GUT theory is. A necessary condition is that $N_f$ falls within the window
given by \begin{equation}
\frac{N_{c}}{k}<N_{f}< 
\frac{N_{c}}{k-\frac{1}{2}}\, ,\end{equation}
where the lower bound comes from the requirement that $\bar{N}_c>0$ and the upper bound is the condition that 
$\bar{b}_0 < 0$. 
This gives us a strong constraint, since we must have $N_{c}\geq k(2k-1)$.
If for example $k=2$, then the minimal case is 
\be
N_f=6 \qquad
\begin{array}{lll}
 \mbox{elec: }SU(10) & \rightarrow & SU(5)\times SU(5)\times U(1),\nonumber \\
\mbox{mag: }SU(2) & \rightarrow & U(1)^{2}.
\end{array}
\ee
The first case with at least three different group factors in the
magnetic theory (i.e. the first non-trivial unification) is
\be
N_f=10\qquad
\begin{array}{lll}
\mbox{elec: }SU(15) & \rightarrow & SU(8)\times SU(7)\times U(1),\nonumber \\
\mbox{mag: }SU(5) & \rightarrow & SU(2)\times SU(3)\times U(1),
\end{array}\ee
however in this case the matching of the $U(1)$'s is less clear because
the unbroken magnetic theory has vanishing $\beta$-function ($2\bar{N}_{c}=N_{f}$).
The first unambiguous case is 
\be
N_f=11\qquad 
\boxed{
\begin{array}{lll}
\mbox{elec: }SU(17) & \rightarrow & {SU(9)}\times SU(8)\times U(1),\nonumber \\
\mbox{mag: }SU(5) & \rightarrow & SU(2)\times SU(3)\times U(1).
\end{array}}
\ee

Now let us consider the different scales. We take the GUT scale
 $M_{GUT} >\Lambda$ to ensure that
the electric theory unifies in the perturbative (weak coupling) regime.
There are then two possible orderings of the dynamical scales of $SU(N_c)$ and $SU(\bar{N}_c)$
consistent with the matching condition: either $\bar{\Lambda} < \Lambda < \mu$ or 
$\bar{\Lambda} > \Lambda > \mu$. These arise as follows: we have $b_0>0$ and $\bar{b}_0<0$ and 
also $|\bar{b}_0|<|{b}_0| $, and therefore the matching condition \eqref{matching2}
leads to the two situations shown in figure 4. (Similar plots hold for the $SU(r_i)$ and $SU(\bar{r}_i)$ 
constituent factors, with the replacements $\Lambda \rightarrow \Lambda_i$ and $\mu\rightarrow \mu_i\sim \hat{\mu}$.) 
\begin{figure}
\begin{centering}
\includegraphics[bb=30bp 500bp 812bp 850bp,clip,scale=0.7]{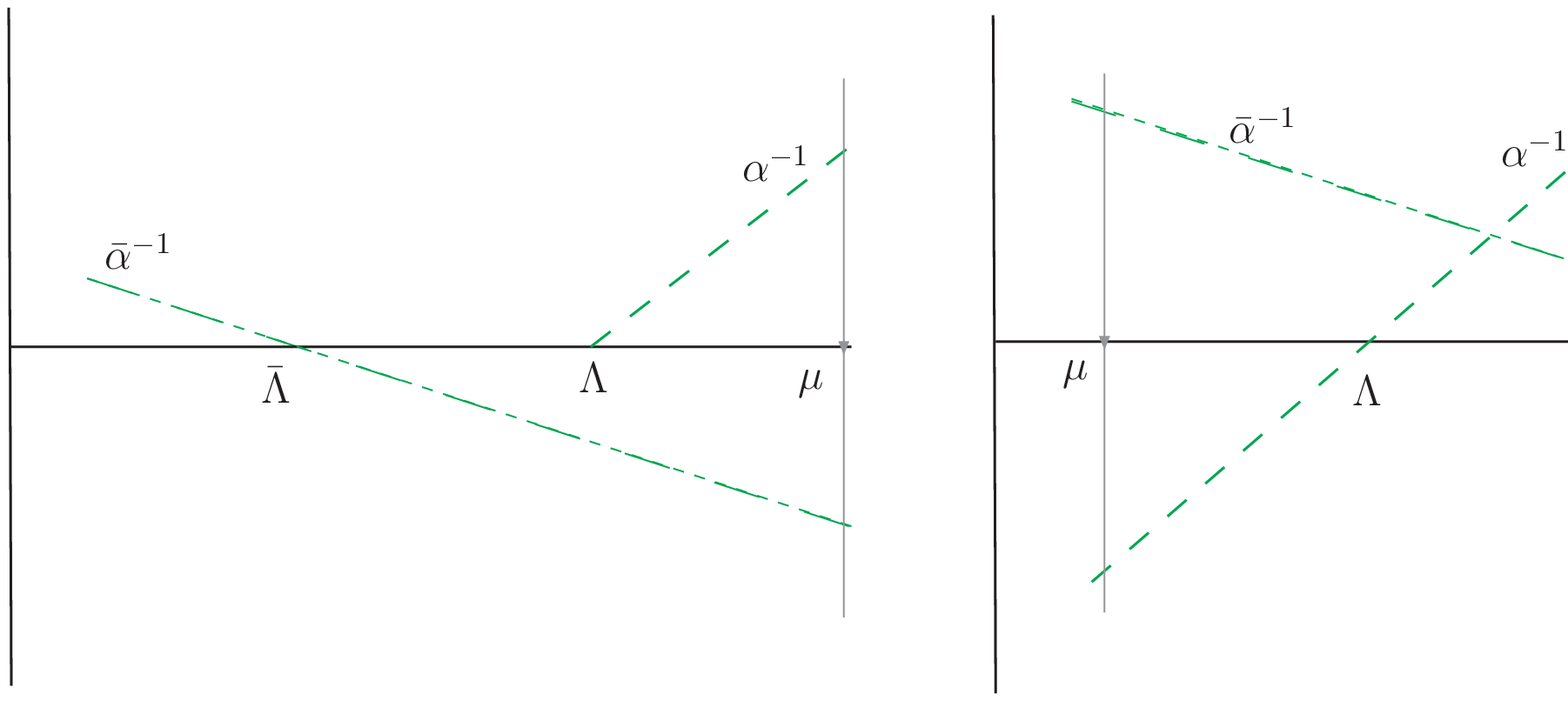}
\par\end{centering}
\caption{\it Running inverse couplings of the unbroken electric $SU(N_c)$ and magnetic $SU(\bar{N}_c)$ 
theories shown as dashed and dot-dashed respectively. The two figures correspond to the two possible orderings 
$\bar{\Lambda} < \Lambda < \mu$ and $\bar{\Lambda} > \Lambda > \mu$.}
\end{figure}

For the first case,
\be
\bar{\Lambda} < \Lambda < \mu,
\ee
the magnetic theory experiences a fake unification below the horizontal axis, 
but the overall magnetic $SU(5)$ theory is never realised as a perturbative theory. 
An example is depicted in figure 5 for the case
 \be
N_f=13\qquad 
\boxed{
\begin{array}{lll}
\mbox{elec: }SU(21) & \rightarrow & SU(11)\times SU(10)\times U(1),\nonumber \\
\mbox{mag: }SU(5) & \rightarrow & SU(2)\times SU(3)\times U(1),\end{array}}
\ee
where we have (rather fancifully) taken $M_{GUT}=2\times10^{16}$GeV.
We also show there
the scales $\mu$ where $\ialpha(\mu)=-\bar{\alpha}^{-1}(\mu)$ in the
unbroken theories, and the scale $\hat{\mu}=\mu^{2}/M_{GUT}$ where
$\ialpha_{i}(\hat{\mu})=-\bar{\alpha}^{-1}_{i}(\hat{\mu})$. 
Note that the unbroken theory 
has a gap where no perturbative description exists and the two theories have no overlap. 
The weak coupling magnetic
description does exist however for the $SU(3) \times SU(2) \times U(1)$ subgroups.

\begin{figure}
\begin{centering}
\includegraphics{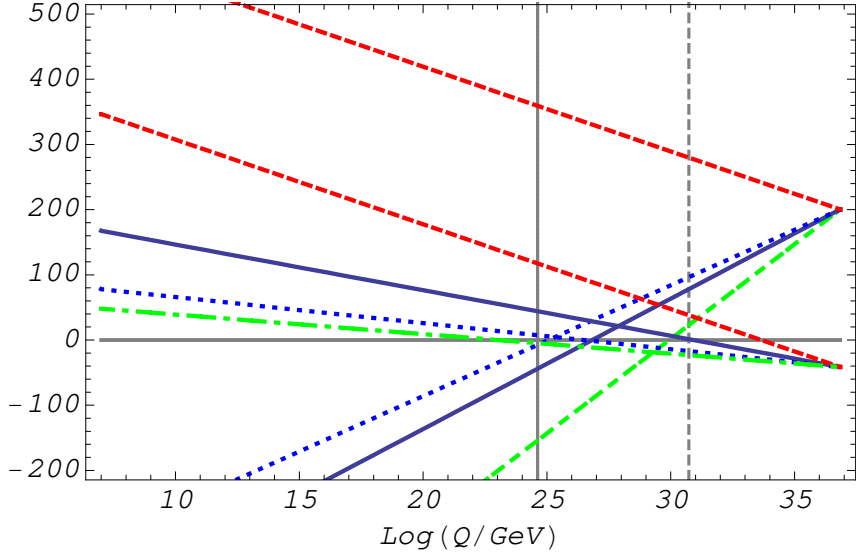}
\par\end{centering}
\caption{\it Running inverse couplings in KSS models with broken GUTs with $M_{GUT},\,\mu>\Lambda>\bar{\Lambda}$ and $k=2$ and assuming $t_0=1$. The couplings
are $U(1)\equiv\mbox{red/dashed}$; $SU(11)\rightarrow SU(2)\equiv\mbox{blue/dotted}$;
$SU(10)\rightarrow SU(3)\equiv\mbox{dark-blue/solid}$. We also show the
running (in green) of the unbroken theory, the scale $\hat{\mu}=\mu^{2}/M_{GUT}$
in solid grey, and the scale $\mu$ in dashed grey. The couplings of the 
unbroken theories obey $\bar{\alpha}({\mu})^{-1}=-{\alpha}({\mu})^{-1}$, while those of the 
$SU(r_i)$ subgroups in 
the broken theories obey $\bar{\alpha}(\hat{\mu})^{-1}=-{\alpha}(\hat{\mu})^{-1}$. For this choice of parameters the 
unbroken theories have no overlap, but the broken theories do.}
\end{figure}

With the complimentary ordering of scales, 
\be
\mu < \Lambda < \bar{\Lambda},
\ee
there are two possibilities: $\Lambda < M_{GUT} < \bar{\Lambda},$ or $\bar{\Lambda} < M_{GUT}.$
In the first case, the magnetic theory also undergoes normal perturbative unification 
(i.e. at positive values of the {\em magnetic} coupling constants). 
However, in the second case, the magnetic theory exhibits a fake unification at negative $\bar{\alpha}.$ Examples are shown in figures 6 and 7.

\begin{figure}
\begin{centering}
\includegraphics{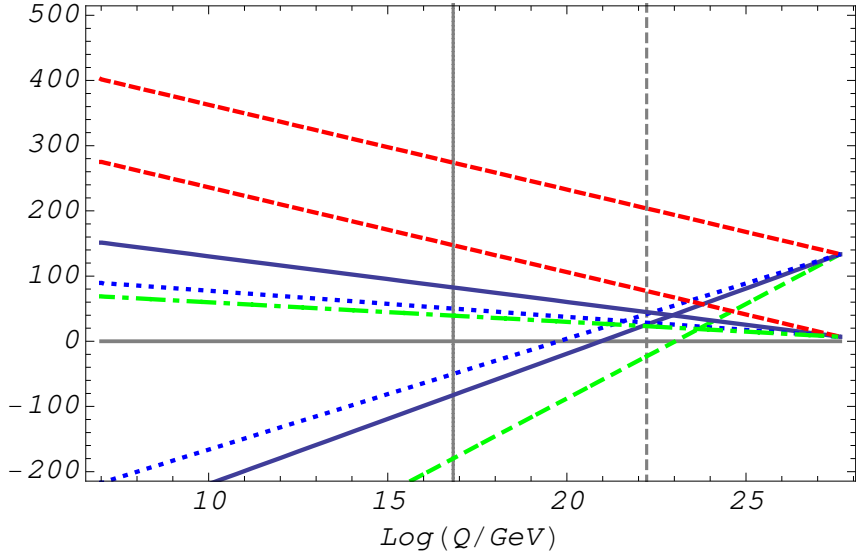}
\par\end{centering}
\caption{\it As in figure 5, for $\bar{\Lambda}>M_{GUT}>\Lambda>\mu $.}
\end{figure}
\begin{figure}
\begin{centering}
\includegraphics{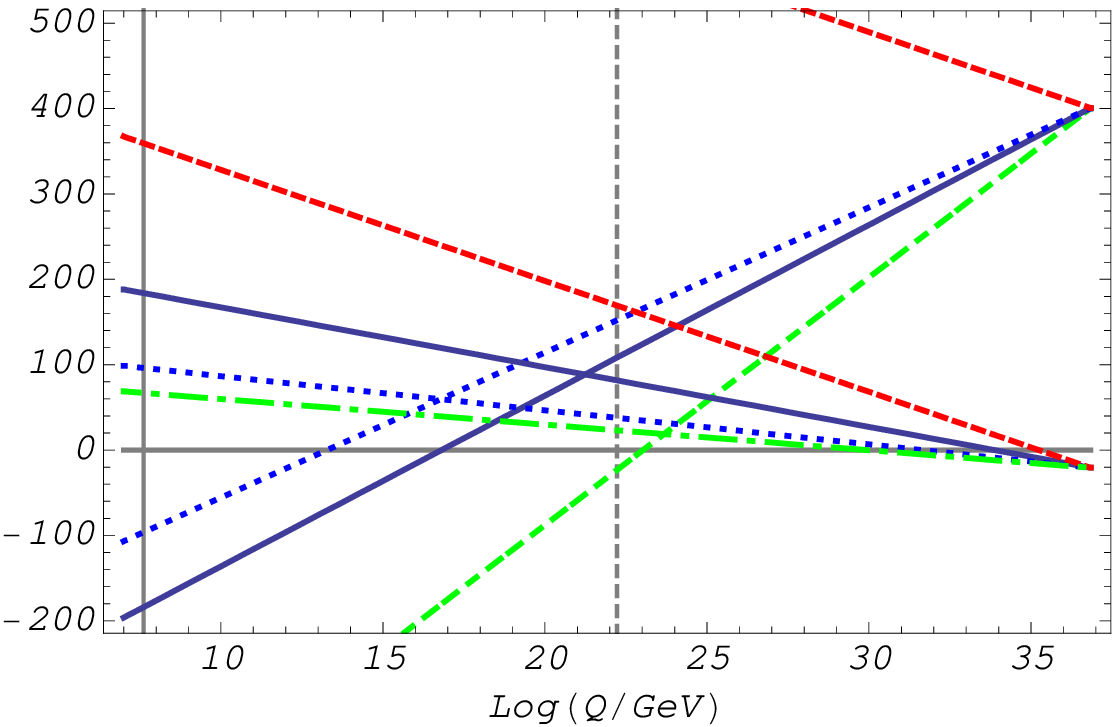}
\par\end{centering}
\caption{\it As in figure 5, for $M_{GUT}>\bar{\Lambda}>\Lambda>\mu $. For this choice of parameters 
the broken and unbroken theories all have perturbative overlap and the magnetic theory unifies at unphysical values.}
\end{figure}

The figures
highlight a few important features. First, even when $\mu\sim\Lambda\sim\bar{\Lambda}$
in the unbroken theory, $\hat{\mu}$ is very different
from the $\Lambda_{i}$ and $\bar{\Lambda}_{i}$. Thus in the broken
theory the dynamical scales are spread by the GUT breaking, and the 
broken theory enjoys a much larger overlap between the electric and 
magnetic descriptions than the unbroken theory. Second
the couplings of the broken theory are always above the unbroken ones
in both the magnetic and electric theories (since in both cases we
have lost some adjoints). Hence the condition $\bar{\Lambda}>\Lambda$
which ensures perturbative overlap in the unbroken theory, is sufficient
to ensure $\bar{\Lambda}_{i}>\Lambda_{i}$ for all the SQCD factors
in the broken theory as well. 

\section{More general models (with coupled sectors)}

The KSS models discussed so far were characterized in the IR by a magnetic theory 
broken into completely decoupled SQCD factors. The unification in both the electric and magnetic descriptions was ensured by the matching relation between the dynamical scales of the two dual theories. 
But does dual-unification apply
in more complicated theories, in particular those with coupled SQCD factors?
We now show that it does; the arguments of the previous section can be made completely
general, and are in fact independent of the theory in question relying
on only a few key assumptions about how the dynamical scales in the
electric and magnetic theories are related. 

In order to do this, it is useful to have a working example. We will use the first of
the more general set of models to be found in refs.\cite{brodie,brodie2}.
This is an extension of the KSS models whose electric
theory has $N_{f}$ flavours and {\em two} adjoint fields $X_{1}$ and $X_{2}$
and electric gauge group $SU(N_{c})$. We will repeat the argument
for these models, step by step. The magnetic theory is an $SU(\bar{N}_c)$ model, where 
\be 
\bar{N}_c=3kN_{f}-N_{c}\, .
\ee
We do not need to go into the details of these models, but
can make do with presenting the undeformed electric and magnetic superpotentials:
\begin{eqnarray}
W_{el} & = & s_{0}\frac{X_{1}^{k+1}}{k+1}+s' _0 X_{1}X_{2}^{2}
\nonumber \\
W_{mag} & = & \bar{s}_{0}\frac{Y_{1}^{k+1}}{k+1}+\bar{s}'_0 Y_{1}Y_{2}^{2}+\frac{\bar{s}_{0}\bar{s}'_{0}}{\mu^{4}}\sum_{i=1}^{k}\sum_{j=1}^{3}M_{ij}\tilde{q}Y_{1}^{k-i}Y_{2}^{3-j}q
\label{eq:41}
\end{eqnarray}
where again the tracelessness of $X_{1,2}$ and $Y_{1,2}$ can be enforced by Lagrange multiplier terms, 
and $M_{ij}$ are magnetic mesons, \begin{equation} 
M_{ij}=\tilde{Q}\, X_{1}^{i-1}X_{2}^{j-1}Q.\end{equation}

Note that $s_{0}$ has mass dimension $2-k$ and $M_{ij}$ has dimension
$i+j$, hence the need for the $\mu^{4}$ in the denominator of the last term of eq.\eqref{eq:41}. We can
perform an overall rescaling of the adjoint fields. The parameter
$\mu$ is then fixed but since it depends on the nonholomorphic part
of the theory we have no control over it.

Now we can again consider adding a deformation which gives VEVs of order $M_{GUT}$
to the adjoint fields. Here we can choose the normalization of the adjoints to be
canonical in the electric theory, but in the magnetic theory we can choose it such that the
the couplings $\bar{t}_i$ in the shifted basis of fields match those of the electric theory. By contrast the quarks can be chosen to have canonical normalization in both theories. 
In the above convention, the deformation induces a VEV structure in the magnetic theory that matches that of the 
electric theory, in the same manner as in
the simpler models of ref.\cite{KS2}.
 
For concreteness we will follow the simple explicit breaking pattern discussed
in sections 4.2 and 4.3 of ref.\cite{brodie} and 5.2 of ref.\cite{brodie2}.
Calling $N_{c}=2n+km$, in that example the GUT symmetry for generic $k$ and $m$ can be broken as 
\be
\begin{array}{lll}
\mbox{elec:}\,\,\,\,\,\,\,\qquad SU(2n+km) & \rightarrow & SU(n)\times SU(n)\times SU(m)^{k}\times U(1)^{k+1} \\
\mbox{mag:}\,\,\, SU(3kN_{f}-2n-km) & \rightarrow & SU(kN_{f}-n)\times SU(kN_{f}-n)\times  SU(N_{f}-m)^{k}\times U(1)^{k+1}.
\end{array}
\label{eq:brodie}
\ee
The fields $X_{1,2}$ decompose into adjoints of $SU(n)$'s and $SU(m)$'s  
plus fields $F=(n,\bar{n})$ in the bifundamental representation of $SU(n)\times SU(n)$ and their
conjugates $\tilde{F}$. The $SU(m)$ adjoints get masses of 
order $M_{GUT}$ but the $SU(n)$ adjoints and bifundamental fields remain light.

Let us now match the running in the two theories, keeping the notation
as general as possible. First we need the scale matching of the unbroken
theories. This was determined in ref.\cite{brodie} to be \begin{equation}
\label{brodieeq}
\Lambda^{N_{c}-N_{f}}\bar{\Lambda}^{\bar{N}_{c}-\bar{N}_{f}}=Cs_{0}^{-3N_{f}}(s'_{0})^{-3kN_{f}}\mu^{4N_{f}}.\end{equation}
We can normalize this relation to the scale $M_{GUT}$, by writing
\begin{equation}
b_0t_{\Lambda}+\bar{b}_0t_{\bar{\Lambda}}=(b_0+\bar{b}_0)t_{\hat{\mu}},\label{eq:ass1}\end{equation}
where $\hat{\mu}$ is a scale which can be determined in terms of
$\mu$. On general grounds the matching relation will aways be of this 
form with different functions $\hat{\mu}$ \cite{KS2}. In this particular
example we have $b_0+\bar{b}_0=3kN_{f}$ and (setting for convenience
$C=1$) \begin{equation}
\frac{\hat{\mu}}{M_{GUT}}=s_{0}^{\prime\, -1}\left(\frac{\mu}{M_{GUT}}\right)^{\frac{4}{3k}}\left(\frac{s_{0}}{M_{GUT}^{2-k}}\right)^{-1/k},\end{equation}
but its actual value isn't important for us.
Recall that the next step was to determine the dynamical scales of the subfactors
in terms of those of the GUT theory. This is done by integrating out
the states that are massive; as we have, said electric unification requires that
these states all have mass terms of order $M_{GUT}$ in the holomorphic superpotential
so that we can write \begin{equation}
e^{-\frac{8\pi^{2}}{g^{2}(M_{GUT})}}=\left(\frac{\Lambda}{M_{GUT}}\right)^{b_0}=\left(\frac{\Lambda_{i}}{M_{GUT}}\right)^{b_{i}}.\end{equation}
In the model of ref.\cite{brodie} this includes both the vector bosons
and adjoint fields. 

Likewise the magnetic theory has\begin{equation}
\label{wilson2}
\left(\frac{\bar{\Lambda}}{M_{GUT}}\right)^{\bar{b}_0}=\left(\frac{\bar{\Lambda}_{i}}{M_{GUT}}\right)^{\bar{b}_{i}}.\end{equation}
This is a Wilsonian relation in which the 
kinetic terms are not necessarily canonically normalized. (This was also important
in ref.\cite{KS2} where the independent check of the couplings in
the magnetic superpotential rested on this procedure.) Thus one has to keep in mind that the matching relation 
involves dynamical scales defined in a possibly unphysical renormalization scheme; we shall return to this 
point momentarily.  Multiplying the two we find \begin{equation}
\Lambda_{i}^{b_{i}}\bar{\Lambda}_{i}^{\bar{b}_{i}}=\Lambda^{b_0}\bar{\Lambda}^{\bar{b}_0}M_{GUT}^{b_{i}+\bar{b}_{i}-b_0-\bar{b}_0}.\end{equation}
 We may now convert this into a relation between couplings
at the scale $t\equiv\log E$. Namely, the above gives \begin{equation}
b_{i}t_{\Lambda_{i}}+\bar{b}_{i}t_{\bar{\Lambda}_{i}}=(b_0+\bar{b}_0)(t_{\hat{\mu}}-t_{GUT})+(b_{i}+\bar{b}_{i})t_{GUT},\end{equation}
and hence \begin{eqnarray}
\bar{\alpha}^{-1}_{i}(t) & = & \bar{b}_{i}(t-t_{\bar{\Lambda}_{i}})\nonumber \\
 & = & \bar{b}_{i}(t-t_{GUT})+\bar{b}_{i}(t_{GUT}-t_{\bar{\Lambda}_{i}})\nonumber \\
 & = & \bar{b}_{i}(t-t_{GUT})-b_{i}(t_{GUT}-t_{\Lambda_{i}})-(b_0+\bar{b}_0)(t_{\hat{\mu}}-t_{GUT})\nonumber \\
 & = & \bar{b}_{i}(t-t_{GUT})+\bar{\alpha}^{-1}_{GUT}\end{eqnarray}
where \begin{equation}
\bar{\alpha}^{-1}_{GUT}=-\alpha^{-1}_{GUT}+(b_0+\bar{b}_0)(t_{GUT}-t_{\hat{\mu}}).\end{equation}
Finally again we have the relation for the $U(1)$ factors which is
found by matching the broken to the unbroken magnetic theories and then 
using eq.\eqref{brodieeq}:
\begin{eqnarray}
\left(\frac{\bar{\Lambda}_{U(1)}}{M_{GUT}}\right)^{\bar{b}_{U(1)}} & = & \left(\frac{\bar{\Lambda}}{M_{GUT}}\right)^{\bar{b}_0}.\nonumber \\
 & = & \left(\frac{\Lambda}{M_{GUT}}\right)^{-b_0}\left(\frac{\hat{\mu}}{M_{GUT}}\right)^{b_0+\bar{b}_0}\end{eqnarray}
which gives precisely\begin{equation}
\bar{\alpha}^{-1}_{GUT}=-\ialpha_{GUT}+(b_0+\bar{b}_0)(t_{GUT}-t_{\hat{\mu}}).\end{equation}
Note that the $\beta$-functions $b_i$ and $\bar{b}_i$, and the 
the precise form of $\hat{\mu}$ were not required; the discussion 
would look the same for any pair of electric/magnetic dual GUTs, provided that the matching 
of the unified theories is of the form \eqref{eq:ass1}.

Now let us return to the issue of the normalization. The unification 
we have derived here is in a basis where the adjoints of the magnetic theory 
are not necessarily canonically normalized\footnote{Recall that this applies only to 
the adjoint fields since we had to match their VEVs
in the magnetic theory to those in the electric one.}. Effectively we are using an unphysical 
renormalization scheme, in which the masses in for example eq.\eqref{wilson2}
could be different from the physical ones. Transferring to a canonically 
normalized basis would rescale the $\bar{\Lambda}$'s by the appropriate 
factors, corresponding to threshold corrections in the gauge 
couplings. 

In order to maintain precise unification (i.e. with with the total absence of threshold corrections) 
therefore one has to make the additional assumption either that the normalization of the light states 
is arranged in complete GUT multiplets or that it is degenerate. For example, in the KSS model, the 
unification in the magnetic theory is guaranteed because the quark normalization can 
be canonical when the theories are matched and all of the adjoints are integrated out in complete
$SU(N_c)$ multiplets. In the extended models of refs.\cite{brodie,brodie2} however, the light $F$ and $\tilde{F}$ states
are not in complete $SU(\bar{N}_c)$ multiplets and the complementary states which 
were integrated out weren't either. Hence one gets perfect one-loop unification only when 
one also assumes degenerate masses $M_{GUT}$ for the latter. Here one can expect
threshold corrections to the magnetic (fake) unification.  (Note however that in the 
Standard Model the matter multiplets {\em do} fall into complete $SU(5)$ multiplets.)

\section{Remarks on proton decay}

One of the obvious areas where dual-unification may have significant 
impact is in proton decay. As has been widely discussed, this arises in 
GUT theories due to the presence of GUT bosons and heavy coloured triplets.
If one assumes simple unification in the MSSM at the usual scale 
$M_{GUT}\approx 2\times 10^{16}$~GeV, the resulting lifetime of the 
proton is shorter than the present experimental bounds~\cite{decayreview,decayreview2}, and
simple unification seems to be ruled out.
Because the MSSM seems to indicate simple unification, this is something of a conundrum. 
In this section we argue on general grounds that it can be resolved by 
dual-unification.

As we have seen, under reasonable 
assumptions, the apparent simple unification of the MSSM could be 
indicative of it being a magnetic theory with a set of 
fields appearing in complete GUT multiplets that 
drive it to a Landau pole at some intermediate scale.
If this is the case, grand unification 
takes place in an electric dual, and this has the potential 
drastically to alter proton decay because the 
proton is a baryon of the magnetic theory, whereas the baryon number 
violating operators are generated in the electric theory. A comprehensive 
discussion would require a full understanding of the Seiberg duality 
of some appropriate supersymmetric version of the Standard Model  
which is alas unavailable, but we can develop a general argument based on the 
model of refs.\cite{brodie,brodie2}, by considering an analogous decay.

In order to do this let us first recap the usual proton decay story \cite{decayreview}. 
In {\em non}-supersymmetric $SU(5)$ the proton is able to decay because
$A^{(X)}$ and $A^{(Y)}$ gauge bosons transform as a $(\bar{3},2)$ of the $SU(3)_c\times SU(2)_L$. Collecting them into an $SU(2)_L$ doublet, $A^{(X)}_{ia}$ where $a$ are $SU(2)$ indices and $i$ are $SU(3)_c$ indices, 
the offending terms in the Lagrangian are of the form 
\begin{eqnarray}
{\cal L}_{A^{(X)},A^{(Y)}}&=& \frac{ig}{\sqrt{2}} (A^\mu_{IK}\bar{X}_{JI}\gamma_\mu X_{KJ}+
 A^\mu_{IK}\bar{Q}_{I}\gamma_\mu Q_{K})\nonumber \\ &\supset &
\frac{ig}{\sqrt{2}} A^{(X)\mu}_{i{a}}
(  \varepsilon_{ijk} \bar{u}^c_{k} \gamma_\mu q_{ja}+ 
\bar{q}_{ib}\gamma_\mu e^+_{ab}+ 
\bar{d}_{i}\gamma_\mu l_{a})
\end{eqnarray}
where $e^+_{ab}=e^+ \varepsilon_{ab}$ is an antisymmetric singlet 
of $SU(2)_L$ which comes from 
the antisymmetric $10$ of $SU(5)$.  For the moment we are using the 
usual nomenclature of the MSSM - thus the right-handed fields are denoted $u^c$ 
and $d^c$, $e^c$, and the left-handed doublets $q$ and $l$. 
So integrating out $A^{(X)}_\mu$ generates a term 
\begin{equation}
{\cal L}_{eff} \supset \frac{g^2}{2M_{GUT}^2} \varepsilon_{ijk}\varepsilon_{ab}( \bar{q}_{aj}  \gamma_\mu 
{u}^c_{k} ) 
(\bar{q}_{ib}\gamma_\mu e^+)\, .
\end{equation}
Note that the effective operator is a baryon of $SU(3)$ (and also a baryon of $SU(2)$). Indeed the new operators, since they must violate baryon number but also respect gauge invariance, can only be baryons. The nett 
result is that the proton can decay via processes such as 
$p\rightarrow \pi^0 e^+ $ as in figure 8a.
\begin{figure}
\begin{centering}
\includegraphics[scale=0.6]{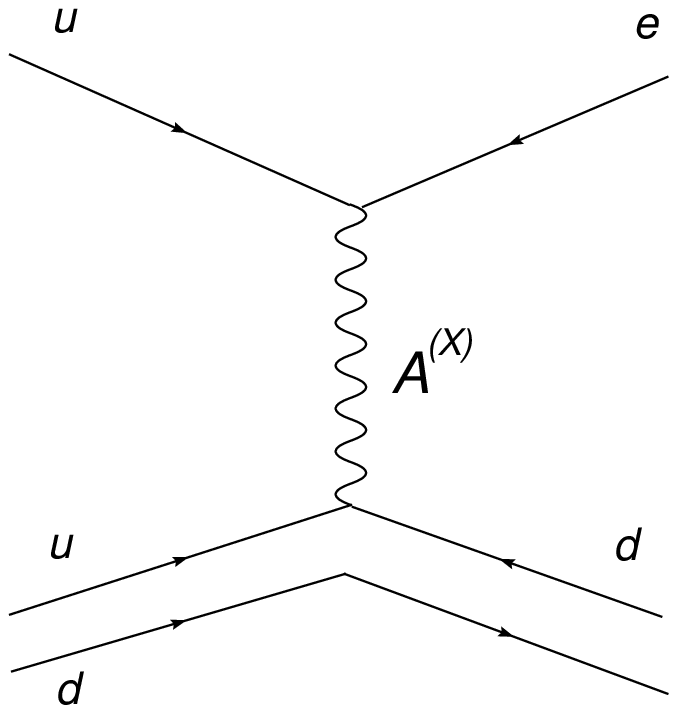}
\hspace{1cm}
\includegraphics[scale=0.7]{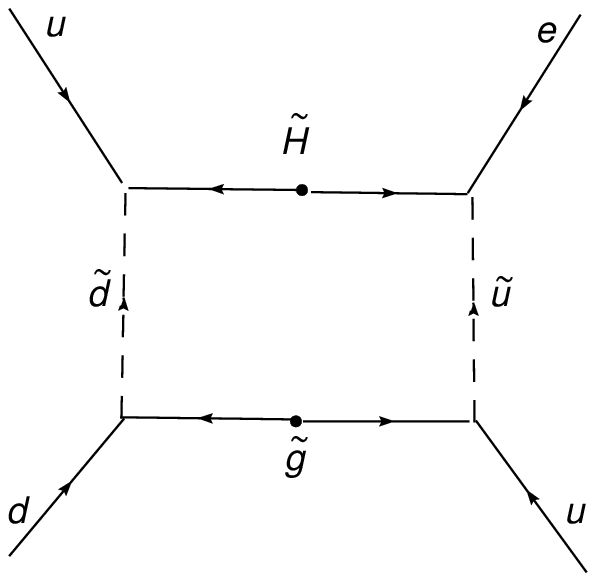}
\par\end{centering}
\caption{\it Proton decay in simple $SU(5)$ SUSY GUTs generated by dimension 6 and dimension 5 operators respectively.}
\end{figure}
These are the dimension 6 operators which exists in 
$SU(5)$ unification. In supersymmetric theories one also has dimension 5 
operators that contribute at one-loop due to the presence of Higgs 
triplets, $\tilde{Q}_T\equiv {\bf \bar{3}}$ and $Q_T\equiv {\bf 3}$, 
that couple via the Yukawa couplings of the MSSM:
\be 
W\supset \frac{h_u}{4} 
\varepsilon_{IJKLM} X_{IJ}X_{KL} H_M + h_d X_{IJ} Q_I \tilde{H}_J 
    \supset 
h_u U^c_i E^c { Q}_{T\,i} +
h_d \varepsilon_{ijk} U^c_i D^c_j {\tilde Q}_{T\,k} \, ,
\ee
and similar for left handed fields.
These give rise via figure 8b to the most dangerous operators; for example 
those involving just the right handed fields are of the form
\be
{\cal L}_{eff} \supset \frac{g^2 h_u h_d}{16 \pi^2 M_{SUSY}  M_{GUT}} 
\varepsilon_{ijk}( {u}^c_{i} e^c ) ( u^c_{j} d^c_{k} )  \, .
\ee
where $h_u$ and $h_d$ are the Yukawa couplings of the MSSM. 
Note that in this estimate, thanks to the non-renormalization theorem, 
the one loop integral is dominated by the low momentum region 
$k\lesssim M_{SUSY}$, and so $M_{SUSY}$ appears in the denominator. 
In the low energy limit the diagram is equivalent to first evaluating 
the non-renormalizable terms in an effective theory,
\be
\label{weff}
W_{eff} \supset \frac{h_u h_d}{M_{GUT}} 
\varepsilon_{ijk}(  E^c {U}^c_{i} U^c_{j} D^c_{k} )  \, ,
\ee
and then computing the diagram in figure 9 with its corresponding 4-point 
vertex.

\begin{figure}
\begin{centering}
\includegraphics[scale=0.8]{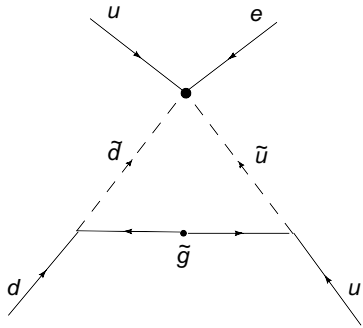}
\par\end{centering}
\caption{\it Approximation to figure 8b in which the dimension 5 operator is evaluated in the electric theory.}
\end{figure}

In a dual-unified theory however, although the magnetic theory appears to 
be unified, proton decay has to go through the electric theory since that is 
where the vector fields and Higgs triplets gain their mass. At this energy 
scale the magnetic theory is strongly coupled and one must instead use the 
weakly coupled electric theory description. In principle 
this could always be done by using the diagram in figure 9. One would first 
compute the relevant operator in the electric theory 
and then map it to the corresponding operators in $W_{eff}$ 
of the magnetic theory via Seiberg duality. 

If one can find the electric dual of the SSM and its GUT theory, 
one has a ready mapping between 
the baryonic operators involved. Since we do not yet know of a such 
theory, we will present a general argument for what happens, 
and then support it by examining an analogous process in a theory where 
both the dual theories are known, namely that of the previous 
section~\cite{brodie,brodie2}. 

First the general argument. Suppose that $SU(3)_c$ baryons of the SSM are
mapped to baryons of $SU(N_c)$ in the electric dual. Our generic picture is 
that the $SU(3)_c$ group factor is strongly coupled in the UV above the messenger scale and 
the $SU(N_c)$ factor is asymptotically free. Hence $N_c > 3$, and 
as we have seen it is typically much larger. Therefore the baryon in the 
electric theory into which $W_{eff}$ maps will have dimension $>4$; 
let us call this dimension $d$, so that schematically the baryon mapping 
would be 
\be 
\varepsilon _{ijk} 
E^c {U}^c_{i} U^c_{j} D^c_{k} \rightarrow \Lambda^{4-d} \chi^d\, ,
\ee
where $\chi $ represents generic fields of the electric theory.
(For convenience we are setting the dynamical scales $\Lambda $ and 
$\bar{\Lambda}$ to be equal.) Now we must look to the electric theory 
to generate the operator in an honest perturbative tree-level diagram 
involving propagators with $M_{GUT}$ scale masses. 
On dimensional grounds we will find  
\be 
W_{el} \supset \frac{\chi^d}{M_{GUT}^{d-3}}\, .
\ee
Note that this is the largest such an operator could be. In principle 
the operator could be smaller if non-renormalizable Planck suppressed 
operators are involved (in which case powers of $M_{Pl}^{-1}$ would 
have to be accommodated as well). The relevant baryon number violating
operator induced in the effective magnetic theory would then be 
\be 
W_{eff} \supset \left( \frac{\Lambda}{ M_{GUT}} \right)^{d-4} 
\frac{1}{M_{GUT}} \varepsilon _{ijk} 
E^c {U}^c_{i} U^c_{j} D^c_{k}  \, .
\ee
Hence the proton decay gets an extra 
$\left( \frac{\Lambda}{ M_{GUT}} \right)^{d-4}$ suppression 
compared with \eqref{weff}, which for even modestly small 
$\Lambda$ would make it ineffective.

It is perhaps clearer why this happens if one begins 
by building equivalents to figure 8b in the electric theory.
In order to generate gauge invariant operators, 
all such diagrams would have many more quark legs since 
they have to correspond to baryons of the electric theory. At low 
external momenta these quark legs confine into electric 
baryons, which can then be mapped into magnetic baryons with the 
accompanying supression. (Of course the magnetic $SU(3)_c$ theory 
only becomes confining again well below the messenger scale.)

Now let us show explicitly that this happens in an analogous process. Consider
the two adjoint models of eq.\eqref{eq:brodie}, with $k=4$ and $m=0$ 
in which the broken model is\footnote{Note that refs.~\cite{brodie,brodie2} also considered 
the $SU(n)\times SU(n') $ structure with $n'\neq n$ and also $N'_f\neq N_f$ for which 
electric/magnetic duality was established, but the unification in this case is more obscure.}
\be 
\begin{array}{lll}
\mbox{elec:} \,\,\,  SU(2n)  & \rightarrow &  SU(n)\times SU(n)' \times U(1) \\
\mbox{mag:} \,\,\,   SU(6)  & \rightarrow & SU(3)\times SU(3)' \times U(1)
\end{array}
\ee
where $6N_f-n=3 $.
We use a prime to distinguish the second $SU(n)$ factor; i.e. in the broken theories the 
field content is $N_f$ flavours of quarks and antiquarks (labelled $Q$, $\tilde{Q}$ and $Q'$, $\tilde{Q}'$ in the electric 
theory and $q$, $\tilde{q}$ and $q'$, $\tilde{q}'$ in the magnetic theory), a single 
massless adjoint for each $SU$ factor (labelled $X$, $X'$ in the electric 
theory and $Y$, $Y'$ in the magnetic theory) and a pair of massless 
bifundamentals (labelled $F$, $\tilde{F}$ in the electric 
theory and $f$, $\tilde{f}$ in the magnetic theory). 

Since the models do not contain asymmetric representations we 
have to improvise a little: we will suppose that the operator 
of interest in the low energy theory is 
\be 
\label{weff2}
W_{eff} \supset \frac{\kappa }{M_{GUT}} \varepsilon_{ijk} 
(Y q)_i  q_j q_k \, .
\ee
Here the adjoint, which has zero baryon number, 
has replaced the right handed electron $E^c$, which came from the
antisymmetric in $SU(5)$.
We are interested in estimating the value of the constant $\kappa $.
We require the baryon mappings of the broken theory which may be obtained from 
ref.\cite{brodie}; they involve both the fundamental and the ``dressed'' 
quarks (i.e. quarks multiplied by some combination of adjoints and bifundemantals); 
in the magnetic theory these are labelled 
\begin{eqnarray} 
q_{(l,1)} &=&  Y^{l-1} q \nonumber \\
q_{(l,2)} &=&  Y^{l-1} \tilde{f} q' \nonumber \\
q_{(l,3)} &=&  Y^{l-1} \tilde{f} f q \nonumber \\
q'_{(l,1)} &=&  \yp^{l-1} q' \nonumber \\
q'_{(l,2)} &=&  \yp^{l-1} \tilde{f} q \nonumber \\
q'_{(l,3)} &=&  \yp^{l-1} \tilde{f} f q' \,\,\, ;\,\,\,\, l=1,\ldots \frac{k}{2}=2 
\end{eqnarray}
and similar for the electric theory with the obvious replacements. Thus, dropping the 
$SU(3)$ indices,  our operator can be written
\be 
W_{eff} \supset \frac{\kappa }{M_{GUT}}
q_{(2,1)}  q_{(1,1)} q_{(1,1)} \, .
\ee
The mapping of this baryon to one of the electric theory is \cite{brodie}
\be 
q_{(2,1)}  q_{(1,1)} q_{(1,1)}
\leftrightarrow 
Q^{\prime\, N_f}_{(1,1)}
Q^{\prime\, N_f}_{(1,2)}
Q^{\prime\, (N_f-1)}_{(1,3)}
Q^{\prime\, N_f}_{(2,1)}
Q^{\prime\, N_f}_{(2,2)}
Q^{\prime\, (N_f-2)}_{(2,3)} \, .
\ee
Note that there are $6N_f-3=n$ indices as required for the $SU(n)'$
contraction. This object has dimension $d=15N_f-11$. Thus, if the 
baryon operator is 
perturbatively generated in the electric theory with coefficients of 
order unity, the resulting $W_{eff}$
in \eqref{weff2} has a coupling given by 
\be
\kappa \sim \left( \frac{\Lambda}{ M_{GUT}} \right)^{15(N_f-1)}.
\ee
Since $N_f> 3$ in these models, 
this is miniscule 
for any reasonable $\Lambda/M_{GUT}$.

\section{Conclusions}

We have proposed two ways in which Seiberg duality can save unification when there 
are Landau poles below the GUT scale, in particular in 
models of direct gauge mediation. In ``deflected-unification", 
the hidden sector experiences strong coupling 
and passes to an electric phase in the UV. This occurs for example when one uses the 
the models of Intriligator, Seiberg and Shih \cite{iss} for the hidden sector. 
As a result the effective number of messenger flavours to which the visible sector couples is reduced in the UV, 
thereby postponing (or even removing) the Landau pole of the SSM to beyond the GUT scale, and 
allowing perturbative unification to take place. 

In "dual-unification", the visible sector is itself a magnetic dual. We showed that 
in known examples where an asymptotically free GUT theory has an IR free magnetic dual, 
the magnetic theory exhibits unification at unphysical values of the gauge coupling reflecting the 
real unification in the electric theory. This arises automatically from the matching relations and we argue that 
it is a general phenomenon. Such unphysical unification is characteristic of 
models of direct mediation in which the messengers are in complete GUT (e.g. $SU(5)$) multiplets, and 
we therefore propose that the SSM could be a magnetic dual theory of this kind. 
The most pressing issue for the dual-unification scenario is of course to find a candidate 
electric/magnetic dual pair for the SSM. 

We also saw that dual-unification can explain why Nature seems to favour unification and 
yet the proton does not decay; in dual-unified theories the unification is only apparent; proton decay 
has to go through baryonic operators induced in the superpotential of the 
electric theory which is the appropriate weakly couple description at the GUT scale; these operators must then 
be matched to the corresponding baryons of the magnetic theory where the proton lives, 
and this procedure comes with a large power suppression. 

The novel feature that Seiberg duality brings to these phenomenological questions 
is a nonperturbative change in the number of degrees of freedom. In the case of deflected-unification 
the duality reduces the effective number of messenger {\em flavours} to which the visible sector couples towards the 
UV. This is a nonperturbative effect; in perturbative field theories new degrees of freedom are (almost always) 
integrated in at higher energy scales. On the other hand the suppression of proton decay in the dual-unification
scenario is a result of a huge increase in the number of {\em colours} of the visible sector in the UV.

\subsection*{Acknowledgements}
We are indebted to Sebastian Franco and Joerg Jaeckel 
for discussions and the Aspen Centre 
for Physics where this work was started. We also thank John March-Russell and Graham Ross for 
insightful comments.

\end{document}